\documentclass[a4paper,UKenglish,cleveref, autoref, thm-restate]{lipics-v2021}

\usepackage{caption}
\usepackage{subcaption}

\newtheorem{problem}{Problem}

\title{Trajectory Range Visibility} 

\titlerunning{Trajectory Range Visibility} 

\author{Seyed Mohammad Hussein Kazemi}{School of Computing and Information Systems, The University of Melbourne, Australia}{kazemis@student.unimelb.edu.au}{}{}

\author{Arash Vaezi}{Department of Computer Engineering, Sharif University of Technology, Iran}{avazei@ce.sharif.edu}{}{}

\author{Mohammad Ali Abam}{Department of Computer Engineering, Sharif University of Technology, Iran}{abam@sharif.edu}{}{}%

\author{Mohammad Ghodsi\footnote{M. Ghodsi's research was partially supported by the Institute for Research in Fundamental Sciences (IPM) under grant No: CS1392-2-01.}}{Department of Computer Engineering, Sharif University of Technology, Iran}{ghodsi@sharif.edu}{}{}

\authorrunning{A. Vaezi et. al.} 

\Copyright{Arash Vaezi, Seyed Mohammad Hussein Kazemi, Mohammad Ali Abam, Mohammad Ghodsi} 



\category{} 

\relatedversion{} 

\supplement{}

\EventEditors{}
\EventNoEds{2}
\EventLongTitle{}
\EventShortTitle{}
\EventAcronym{}
\EventYear{2022}
\EventDate{}
\EventLocation{Little Whinging, United Kingdom}
\EventLogo{}
\SeriesVolume{42}
\ArticleNo{12}

\def\VP{\mbox{\it VP}}
\def\VG{\mbox{\it VG}}
\def\TRVP{\mbox{\it TRVP}}
\def\RTRV{\mbox{\it RTRV}}

\def\PNST{\mbox{\it PNST}}
\def\INST{\mbox{\it INST}}
\def\LRTV{\mbox{\it LRTV}}
\def\SPI{\mbox{\it SPI}}
\def\VRT{\mbox{\it VRT}}
\def\CV{\mbox{\it CV}}
\def\P{\mathcal{P}}

\begin{document}
\nolinenumbers

\maketitle
\begin{abstract}

Consider two entities with constant but not necessarily equal velocities, moving on two given piece-wise linear trajectories inside a simple polygon $\P$. The Trajectory Range Visibility problem deals with determining the sub-trajectories on which two entities become visible to each other. A more straightforward decision version of this problem is called Trajectory Visibility, where the trajectories are line segments. The decision version specifies whether the entities can see one another. This version was studied by P. Eades et al. in 2020, where they supposed given constant velocities for the entities. However, the approach presented in this paper supports non-constant complexity trajectories. Furthermore, we report every pair of constant velocities with which the entities can see each other. 
In particular, for every constant velocity of a moving entity, we specify: $(1)$ All visible parts of the other entity's trajectory. $(2)$ All possible constant velocities of the other entity to become visible. 

Regarding line-segment trajectories, we present $\mathcal{O}(n \log n)$ running time algorithm which obtains all pairs of sub-trajectories on which the moving entities become visible to one another, where $n$ is the complexity of $\P$. Regarding the general case, we provide an algorithm with $\mathcal{O}(n \log n + m(\log m + \log n))$ running time, where $m$ indicates the complexity of both trajectories. We offer $\mathcal{O}(\log n)$ query time for line segment trajectories and $\mathcal{O}(\log m + k)$ for the non-constant complexity ones s.t. $k$ is the number of velocity ranges reported in the output. Interestingly, our results require only $\mathcal{O}(n + m)$ space for non-constant complexity trajectories.

\end{abstract}
\section{Introduction}
 
Commonly, trajectory problems relate to classifying or extracting features from a set of trajectories. For example, passing closely together represents an encounter, or moving together for an extended period shows a social
group. K. Buchin et al.~\cite{buchin2012detecting} stated that the movement of monkeys and their behavior changes are points of interest when confronting other social groups. A Brownian bridge is a  model for such cases as in~\cite{buchin2012detecting}. One might then ask: What is the probability of the entities becoming visible? This is answered by K. Buchin et al.~\cite{buchin2019region} for stationary entities whose location is in terms of a probability distribution. Other ongoing research areas are clustering under various distance metrics, finding flocks of entities moving together~\cite{andersson2007reporting,benkert2008reporting}, detecting frequently visited locations a.k.a. hot-spots and inferring road maps based on traffic data (a.k.a. map construction)~\cite{buchin2011detecting,gaffney2007probabilistic}. It is worth mentioning (discrete) Fŕechet distance, informally known as the dog walker’s distance, which is the most commonly used similarity measure between pairs of trajectories. For more information, see~\cite{11475/15060}. 
 
One of the interesting and recent applications of trajectory problems is not in spatial only but also in spatiotemporal data. Moreover, the advent of IoT and the rapid increase in hiring GPS-enabled devices are constantly generating a vast quantity of trajectory data. The main concern here is the entities moving in space over time~\cite{eades2020trajectory}. One can consider such movements as a sequence of \textit{time-stamped points} in $\mathbb{R}^d$, called trajectories.
Thus, demand for a strong theory of trajectories and efficient algorithms naturally escalates. There are more applications of the latter field in, for instance, GIScience, databases, and in relatively further away fields such as meteorology and ecology. Usually, sources for this type of data include the movements of hurricanes~\cite{stohl1998computation}, animals~\cite{calenge2009concept,gurarie2009novel}, traffic~\cite{li2010deriving} and even more~\cite{dodge2009revealing}.
Another study dealt with tracking an entire honey bee colony, including recovering 79 percent of bee trajectories from five observation hives over 5 min timespans~\cite{bozek2021markerless}.
Also, the authors in~\cite{pane2021real} track microrobots in tissues using ultrasound phase analysis. Their work performs real-time tracks of the microrobot's positions along linear trajectories with a linear velocity of up to $1 \ {mm}/s$. Moreover, a UV sensor-based dual-axis solar tracking system, suggested in~\cite{jamroen2021novel}, offers the capability of following the sun's trajectory through daily and elevation angles. An exciting line of research is tracking the location of mobile devices in cellular networks outdoors~\cite{trogh2019outdoor}. The authors collected millions of parallel location estimations from over a million users in Belgium and processed them in real time. Their experiments are conducted with trajectories on foot, by bike, and by car.

\textbf{Trajectories and Visibility.} Consider a trajectory $\tau_q$ with a moving entity $q$, and another trajectory $\tau_r$ with the moving entity $r$ within a simple polygon $\P$.
Let $T_{q,r}$ be the set of time intervals at which $q$ and $r$ can see each other.
The existing visibility tools allow one to check if there are sub-trajectories (at least one point) of $\tau_q$ and $\tau_r$ visible to each other. Note that the entities become visible to one another only if they are in such sub-trajectories. There could be quadratically many pairs of visible sub-trajectories. Nonetheless, the entities might never simultaneously be within such a pair. To determine visibility between moving entities, one needs to incorporate the concept of time into existing visibility tools. The majority of existing research on trajectory visibility concerns kinetic data structures. A data structure representing the visibility polygon of a point $p$ is created and updated as $p$ moves through the environment. The authors of~\cite{aronov2002visibility} have produced research along this line. Results on maintaining the shortest path between two moving entities such as~\cite{diez2017kinetic} can be used to track visibility between two moving entities. A core feature of these kinetic methods is that they are event-based. The time taken to maintain the data structure as a point moves depends on the number of events that occur along the way. This approach is beneficial if the number of events is small, but not when it is proportional to the complexity of the environment. If so, very little is gained by employing such a data structure as computing the polygon from scratch in the new location may be faster and more convenient. A recent work by K. Buchin et al~\cite{buchin2022segment} also exists regarding counting the number of objects visible to a query point inside a simple polygon.
Moreover, Jansen~\cite{jansen2021local} studies the properties of a simple polygon that influence visibility computations on the polygon. On the other hand, considerable research exists on the related problem of trajectory planning under visibility constraints, see~\cite{shkurti2014maximizing}. In their problem, while maximizing the time they are visible, two collaborative robots must move through a terrain subject to constraints on their movement.

Additionally, other tools useful in the context of analyzing trajectory pairs may be as follows: (1) The \textit{Graham algorithm}~\cite{graham1972efficient} that finds the convex hull of a given set of points on a plain. (2) The \textit{endpoint tree}, introduced by Qiao et al.~\cite{qiao2016range}. The endpoint tree aims to efficiently process an element of a provided input stream that might fall into a large set of given ranges. (3) The authors of~\cite{qiao2016range} use a technique to find the \textit{minimal set} (called \textit{participants}) of nodes on an endpoint tree for a given range. Note that the range must be among the set of ranges the endpoint tree is built based on it. The members of this set cover disjoint intervals whose union is equal to the provided range. Please refer to~\cite{qiao2016range} for more details.

\textbf{Previous Results.} Provided the polygon $\P$ with $n$ vertices, P. Eades et al.~\cite{eades2020trajectory} introduce only the case of whether there exists a point at the time where two entities are observable for one another. For this version, the authors in~\cite{eades2020trajectory}  temporally decompose the problem: The answer to the visibility question is \textit{no} if and only if entities remain invisible to each other on every pair of time stamps $t$ and $t^{'}$ s.t. $t^{'} = t + 1$.  Given the above settings, when both entities move along a line segment, the following cases appear in~\cite{eades2020trajectory}: $(a)$ A simple polygon setting. $(b)$ A simple polygon where the entities may move through obstacles. $(c)$ A polygonal domain in which the entities may move through obstacles. Consider a large constant $k$. The algorithmic run-time of $(a)$ would be $\Theta(n)$ while requiring $\mathcal{O}(n \log^5 n)$ for pre-processing, as well as space consumption, and $\mathcal{O}(n^{3/4} \log^3 n)$ as the query time (see~\cite{eades2020trajectory}). On the other hand, the bounds of $(b)$ and $(c)$ are interestingly almost identical: $\mathcal{O}(\log^k n)$ for the query time, $\mathcal{O}(n^{3k})$ as the space complexity and the data structure pre-processing time, and $\mathcal{O}(n \log n)$ for the algorithmic running time of (b). The latter is $\Theta(n \log n)$ when it comes to $(c)$ (again, see~\cite{eades2020trajectory}).


\subsection{Our Contribution.} 

\setlength{\textfloatsep}{10pt}
\setlength{\intextsep}{10pt}
\begin{table}[]
    \caption{\label{tab:ourResults}The two leftmost columns specify if the trajectory $\tau_q$ and $\tau_r$ for moving entities $q$ and $r$ are line segments (/) or sets of line segments, which we represent as a set of vertices ($\cdot$) with the size $m$. Also, by R.S. in column four, we mean ray shooting pre-processing time. The second row shows our results for a restricted version of Problem~\ref{TRSP}. We then present our results for the general form of the same problem in the third row. Note that we always assume that the polygon $\mathcal{P}$ is simple s.t. $n$ is the number of vertices of $\mathcal{P}$. In contrast to the previous works, we can output the range of velocities that make the entities visible to each other. We also denote the number of the endpoints of such ranges as $k$.  }
    \centering
    \setlength{\arrayrulewidth}{0.5mm}
    \setlength{\tabcolsep}{10pt}
    \renewcommand{\arraystretch}{1.1}
    \begin{tabular}[htp]{ |p{0.01cm}|p{0.01cm}|p{3cm}|p{1.5cm}|p{0.9cm}|p{1.8cm}|p{0.8cm}| }
    \hline
    \multicolumn{7}{|c|}{Our Contribution}\\
    \hline
    $\tau_q$ & $\tau_r$ & $\ \ \ $  Algorithm & $\ \ \ $   Space & R.S. & Query & Src. \\
    \hline
    / & / &  $\mathcal{O}(n \log n)$ & $\mathcal{O}(n \log n)$ & $\mathcal{O}(n)$ & $\mathcal{O}(\log n)$ & Sec.~\ref{subsec:FirstVariant} \\
    $\cdot$ &$\cdot$ & $\mathcal{O}(n + m \log nm)$ & $\mathcal{O}(n + m)$ & $\mathcal{O}(n)$ & $\mathcal{O}(\log m + k)$ & Sec.~\ref{nonSegmentRestricted} \\
    $\cdot$    & $\cdot$ & $\mathcal{O}(n \log n + m\log mn)$ & $\mathcal{O}(n + m)$ & $\mathcal{O}(n)$ & $\mathcal{O}( \log m + k)$ &  Sec.~\ref{subsection:NonSegTrajs} \\
    \hline
    \end{tabular} 
    \label{tab:my_label}
\end{table}

In this paper, we obtain the results in Table~\ref{tab:ourResults}, and extend the work of P. Eades et al.~\cite{eades2020trajectory} by considering the following crucial elements in the problem we solve:

    (1) Extending the problem from merely its decision version. So, we examine the trajectories to find \underline{all pairs of} sub-trajectories on which moving entities can see one another.

    (2) Considering the trajectories with \underline{not necessarily constant complexity}. Specifically, assuming each trajectory as a \textit{set of vertices} s.t. a \textit{line segment} can exist between \textit{some pairs} of vertices in that set.
    
    (3) The velocity of the entities \underline{are not necessarily given}. We provide a fast query time for specifying a \textit{range} of velocities, s.t. each value on those ranges implies \textit{different (sets of) sub-trajectories}, on which two entities become visible.
    
In Section~\ref{sec:preliminaries}, we formally define the general variant of the Trajectory Visibility problem that first appeared in~\cite{eades2020trajectory}. Some details regarding the techniques used in this paper will be discussed as well. Next, Section~\ref{sec:completlyVisTrajs} considers a crucial case of the problem, where all points in both trajectories are visible to one another. Illustrating the details of the latter, Section~\ref{sec:givenvelocities} solves the first version of the problem introduced in Section~\ref{subsec::ProblemDefinition}. On the other hand, Section~\ref{sec:arbitraryvelocity} aims to offer the solution for the general case of the problem.
    
Structure-wise, the main results of this paper appear as follows: Specifically for Problem~\ref{TRGV}, Lemma~\ref{lemma:VRT} offers a solution with a small error rate of $\frac{1}{L} \leq \varepsilon$ s.t. $L$ will be defined in Subsection~\ref{subsec::ProblemDefinition}. On the other hand, a simplified variant of Problem~\ref{TRSP} is solved in Theorem~\ref{theorem:lineSegProb2}. However, Theorem~\ref{theorem:INST} would solve the general form of Problem~\ref{TRSP}. In order to prove Theorem~\ref{theorem:INST}, Lemma~\ref{lemma:nonSegPolyTime} and Theorem~\ref{theorem:RestrictedTRSP} before that, appear in this work as well.

\section{Preliminaries}
\label{sec:preliminaries}

We first formally define a general version of the Trajectory Visibility Problem stated in~\cite{eades2020trajectory}. Call the general variant Trajectory Range Visibility ($\TRVP$). Second, we discuss the Visibility Glass ($\VG$) technique that appeared in~\cite{eades2020trajectory}, introduced based on Hourglass~\cite{guibas1989optimal}. These preliminaries will ease the understanding of our proposed techniques. Also, the challenges our algorithm is trying to resolve will be illustrated in this section. 

\subsection{Problem Definition}
\label{subsec::ProblemDefinition}



In our settings, we always assume that $q$ and $r$ cannot see through the edges of $\mathcal{P}$. Also, we consider a trajectory $\tau$ as a set of connected endpoints $V_{\tau}$ (trajectory vertices connected by trajectory edges) contained in a polygon $\mathcal{P}$. Moreover, by visibility, we specifically mean \textit{weak visibility}. Accordingly, we first start with a simpler variant of the problem, then remove some of the constraints and provide a more general solution. The first variant is as follows:

\begin{problem}[Trajectory Range Visibility with Given Velocities.]
\label{TRGV}
    Denote the velocities of $q$ and $r$ provided as $v_q(t) = C_0$ and $v_r(t) = C_1$ s.t. $C_0$ and $C_1$ are constants but not necessarily equal. Let $\tau_q$ and $\tau_r$ be the corresponding trajectories of $q$ and $r$, both given as two distinct line segments inside a simple polygon $\mathcal{P}$. Consider $v_q(t)$ and $v_r(t)$, given on $\tau_q$ and $\tau_r$. Also, let $T_{q,r}$ be the set of time intervals at which $q$ and $r$ can see each other. Accordingly, the problem is to find all time intervals like $[t_i, t_j] \in T_{q,r} \ {s.t.} \ j > i \ \& \ i, j \in \mathbb{N}$, if there is a sub-segment ${\tau}^{'}_{q} \subset {\tau}_q$ that $q$ passes through it during $[t_i, t_j]$, while there is another sub-segment in ${\tau}_r$ that is visible from ${\tau}^{'}_{q}$, called ${\tau}^{'}_{r} \subset {\tau}_r$, s.t. $r$ moves on ${\tau}^{'}_{r}$ exactly during $[t_i, t_j]$. Note that if ${\tau}^{'}_{q}$ and ${\tau}^{'}_{r}$ are visible to each other, then every point in ${\tau}^{'}_{q}$ sees every point in ${\tau}^{'}_{r}$.
\end{problem}

To illustrate our final solution for Problem~\ref{TRGV}, we apply a restriction first to generate a simpler variant. This constraint will be removed later: \noindent \textit{Integrality of Coordinates Assumption.} Bonnet et al.~\cite{bonnet2016approximation} proved that assuming all coordinates of the vertices of the polygon $\mathcal{P}$ (we also consider the trajectory vertices in the context of Problem~\ref{TRGV}) as integer numbers, provided in the binary form, yields a crucial result. That is, not only the distance between all two vertices becomes at least one, but a useful lower bound, called $d$, appears on the distances between all pairs of two objects that do not share a point. Let $M$ be the largest appearing integer in the problem. Also, define \textit{diam}($\mathcal{P}$) as the largest distance between all pairs of two points in $\mathcal{P}$. Observe that \textit{diam}($\mathcal{P}$) $< 2M$. Denote $L = 20M > 10$. Then it holds that $\frac{1}{L} \leq d$. We will use this setting to restrict the problem we defined above. It would let us avoid visibility-related issues mentioned in~\cite{bonnet2016approximation} as \textit{counter examples} for previous work and possibly beyond that. We will use this setting later in Lemma~\ref{lemma:VRT}.

Using the former problem definition under the latter setting, we list five crucial constraints overall, which we will refer to them frequently in this paper:
\begin{enumerate}
   \item Requiring the form of the trajectories as \textit{line-segments} \label{constraint1}
   \item Assuming \textit{no holes} inside the polygon $\mathcal{P}$ \label{constraint2}
   \item Forcing the trajectories to \textit{always remain inside} the polygon \label{constraint3}
   \item Having integer coordinates for the polygon and trajectory vertices \label{constraint4}
   \item Considering $v_q(t)$ and $v_r(t)$ \textit{given} as two constants $C_0$ and $C_1$ \label{constraint5}
\end{enumerate}

We will later introduce a solution that removes the constraints (\ref{constraint4},  \ref{constraint5}) as a first step. Note that we will remove (\ref{constraint5}) by finding a \textit{range} of all values for $C_0$ and $C_1$, maintaining the visibility between $q$ and $r$. Then, we nullify constraint (\ref{constraint1}) by considering a trajectory $\tau$ as a set of trajectory vertices $V_{\tau}$ s.t. $|V_{\tau}| > 2$, and for each $u, \ v \in V_{\tau}$, $u$ and $v$ can be the endpoints of a sub-trajectory $\overline{uv}$, in the form of a line segment. Accordingly, the problem without the above constraints can be defined as follows:

\setlength{\textfloatsep}{10pt}
\setlength{\intextsep}{10pt}
\begin{problem}[The Trajectory Range Visibility Problem - $\TRVP$.]
\label{TRSP}
Given two entities $q$ and $r$ and their corresponding trajectories ($\tau_q$ and $\tau_r$) inside $\mathcal{P}$. Suppose the trajectories are a set of line segments (in contrast to Problem~\ref{TRGV}) connected from their endpoints. Assume the trajectories cannot intersect the edges of $\mathcal{P}$.
Define ${\tau}^{*}_{r}$ to be a sub-trajectory of ${\tau}_{r}$, and  ${\tau}^{*}_{q}$ to be a sub-trajectory of ${\tau}_{q}$. 
Find every time interval $[t_i, t_j] \in T_{q,r} (\ j > i \ \& \ i, j \in \mathbb{N})$ that:

 For all constant values $C_0$ and $C_1$, $q$ moves with the velocity of $v_q(t) = C_0$, and $r$ moves with the velocity of $v_r(t) = C_1$. 
 
  For all the sub-trajectories like ${\tau}^{*}_{q}$, $q$ passes through ${\tau}^{*}_{q}$ during $[t_i, t_j]$, and sub-trajectories like ${\tau}^{*}_{r}$ that $r$ moves on ${\tau}^{*}_{r}$ simultaneously during $[t_i, t_j]$, while $r$ and $q$ can see each other.
\end{problem}

\noindent \textbf{Further Illustrations on Problem~\ref{TRSP}.} The Fréchet distance ($\delta_f$) is developed by french mathematician Maurice Rene Fréchet. Imagine two trajectories $\tau_q$ and $\tau_r$ inside the simple polygon $\mathcal{P}$, consisting of points with straight lines connecting the points. The paths themselves may cross, but $q$ and $r$ may never switch paths, nor may they move backward: they may only move forward or maintain their current position. Between the $q$ and $r$ is a leash. The Fréchet distance then is the minimum length of this leash s.t. $q$ and $r$ may both reach the end of their respective paths. This length is called $\Delta$ and is the solution to the problem. Determining whether $q$ and $r$ can complete their paths, given a leash of length $\Delta$, is referred to as the decision problem. This variant can be solved using a so-called Freespace diagram. A Freespace diagram is a two-dimensional matrix. Given two trajectories $\tau_q$ and $\tau_r$ and a Fréchet distance $\Delta$, the diagram contains each combination of points on the trajectories whether $\delta(q, r) \leq \Delta$. If so, the diagram marks it as free. 

\begin{figure}
     \begin{subfigure}[b]{0.5\textwidth}
        \includegraphics[width=\textwidth]{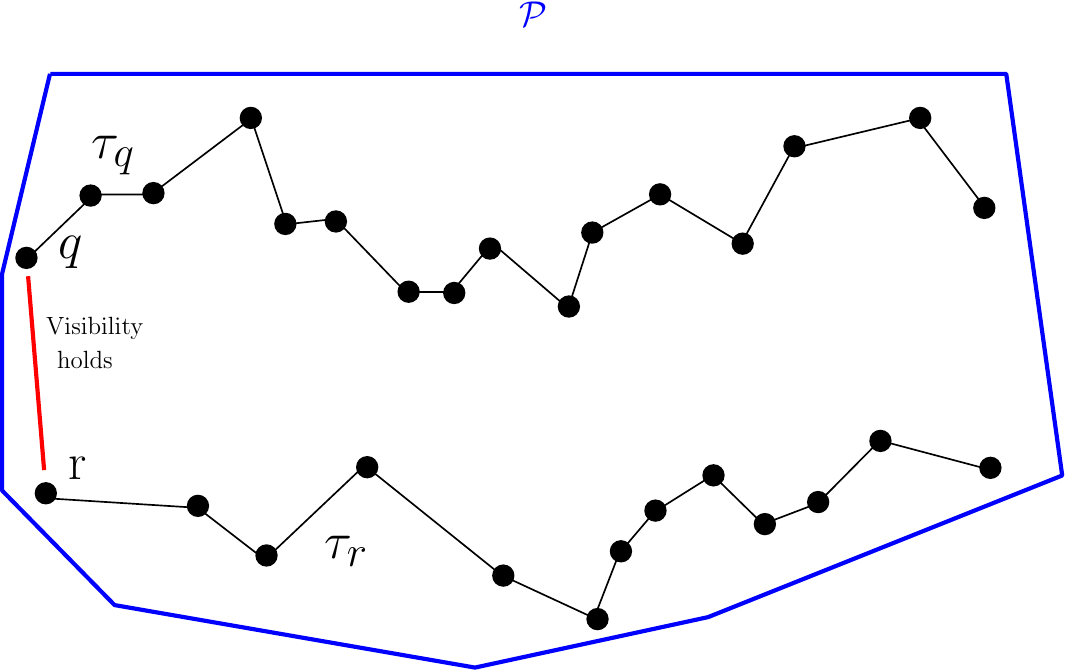}
        \caption{Arbitrarily Picked Trajectories of $q$ and $r$}
        \label{APT}
     \end{subfigure}
     \hfill
     \begin{subfigure}[b]{0.4\textwidth}
         \includegraphics[width=\textwidth]{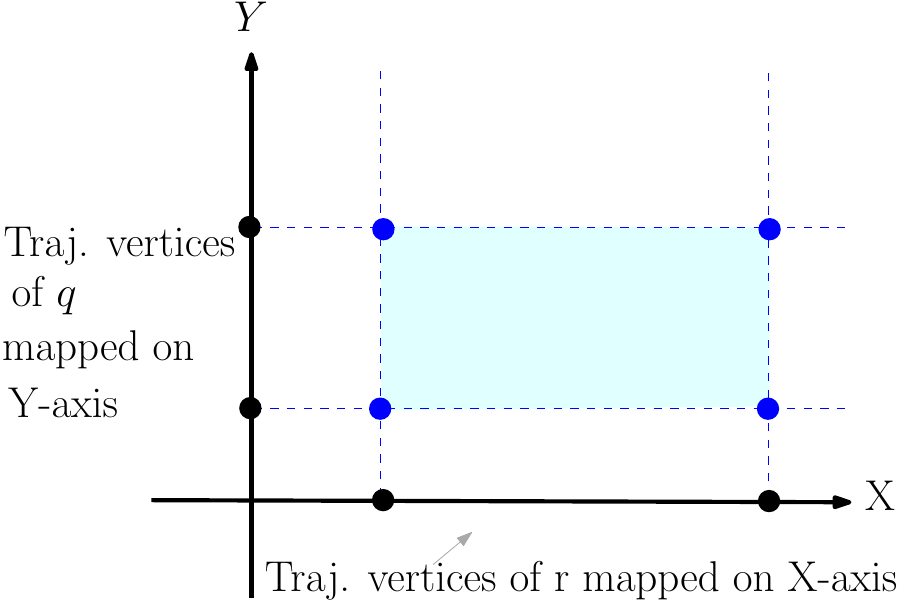}
         \caption{Visibility Diagram}
         \label{VFD}
     \end{subfigure}
     \hfill
     \caption{Based on whether the weak visibility holds between two moving entities $q$ and $r$, Problem~\ref{TRSP} can be explained using a Visibility diagram; in which every point $p_f = (x, y)$ indicates $r$ is at a particular position $p_{\tau_r} \in \tau_r$, whose mapped point on X-axis is $x$, while $r$ sees $q$ at $p_{\tau_q} \in \tau_q$ s.t. the mapping of $p_{\tau_q}$ on Y-axis is $y$. Therefore, Problem~\ref{TRSP}, in this setting, becomes as reporting all such points like $p_f$ considering all constant values for $v_q(t)$ and $v_r(t)$.}
     \label{Diag1}
\end{figure}

\begin{figure}
    \begin{subfigure}[b]{0.5\textwidth}
        \includegraphics[width=\textwidth]{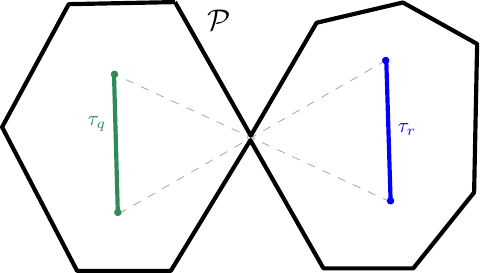}
        \caption{Each point in $\tau_q$ sees one point in $\tau_r$}
        \label{polygonAndTrajs}
     \end{subfigure}
     \hfill
     \begin{subfigure}[b]{0.4\textwidth}
         \includegraphics[width=\textwidth]{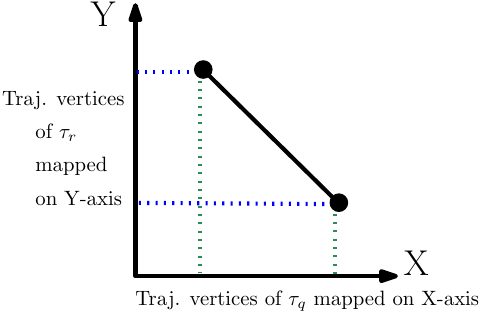}
         \caption{Visibility Diagram for $\tau_q$ \& $\tau_r$}
         \label{visDiagramOfTaus}
     \end{subfigure}
    \hfill
    \caption{Figure~\ref{polygonAndTrajs} shows the polygon $\mathcal{P}$ and two trajectories $\tau_q$ and $\tau_r$ within it. The form of $\mathcal{P}$ is in such a way that every point $x \in \tau_q$ sees exactly one point $y \in \tau_r$. On the other hand, Figure~\ref{visDiagramOfTaus} illustrates the corresponding visibility diagram of $\tau_q$ and $\tau_r$. Notice that the trajectory vertices of $\tau_q$ and $\tau_r$ are mapped on X-axis and Y-axis, respectively.}
    \label{VisDiag}
\end{figure}

In the context of Problem~\ref{TRSP}, the leash shows whether the weak visibility holds between $q$ and $r$. The moving entities may move forward only, with constant velocities $v_q = C_0$ and $v_r = C_1$. Therefore, denote the Visibility diagram as a Freespace diagram that marks each combination of points on the trajectories, whenever the weak visibility holds between $q$ and $r$. See Figures~\ref{Diag1} and~\ref{VisDiag}. Accordingly, Problem~\ref{TRSP} becomes as finding all points that the Visibility diagram marks as free, for all constant values $v_q = C_0$ and $v_r = C_1$.

\begin{observation}
\label{obsFrech}
    Under the constraints~\ref{constraint1},~\ref{constraint2},~\ref{constraint3}, Problem~\ref{TRSP} is solvable by drawing a Visibility diagram $V(\tau_q, \tau_r)$ for trajectories $\tau_q$ and $\tau_r$. Moreover, removing constraint~\ref{constraint1} makes the problem non-trivial.
\end{observation}

\begin{proof}
    Observe that the marked points on $V(\tau_q, \tau_r)$ may form a simple polygon called $\mathcal{P}_v$. Note that, in some cases, the diagram becomes a line segment (see Figure~\ref{VisDiag}). So, if $V(\tau_q, \tau_r)$ forms a polygon, due to its definition, all points inside $\mathcal{P}_v$ are marked. But for convenience, consider the surroundings of $\mathcal{P}_v$ only. Therefore, for a point $x_q \in \tau_q$ (w.l.o.g mapped on the X-axis), denote $\tau^{*}_{r} \subset \tau_r$ as a sub-trajectory visible for $x_q$. Hence $\tau^{*}_{r}$ might be only a point (similar to Figure~\ref{VisDiag}) or empty (which means $x_q$ does not see even one point on $\tau_r$). Yet if $\tau^{*}_{r}$ becomes a line segment, there are two points in which a line parallel to Y-axis and crossing $x_q$ intersects $\mathcal{P}_v$. Thus, two ray-shooting queries on $\mathcal{P}_v$ can determine the latter intersection points. However, by removing constraint~\ref{constraint1}, there may be many more polygons like $\mathcal{P}_v$. This is mainly because of letting the trajectories consist of many line segments. Accordingly, many pairs of line segments from $\tau_q$ and $\tau_r$ respectively, require the formation of a Visibility diagram like $\mathcal{P}_v$. 
\end{proof}

\subsection{Some Notations and Definitions}
\label{subsec::Defs}
\begin{figure}
     \begin{subfigure}[b]{0.5\textwidth}
        \includegraphics[width=\textwidth]{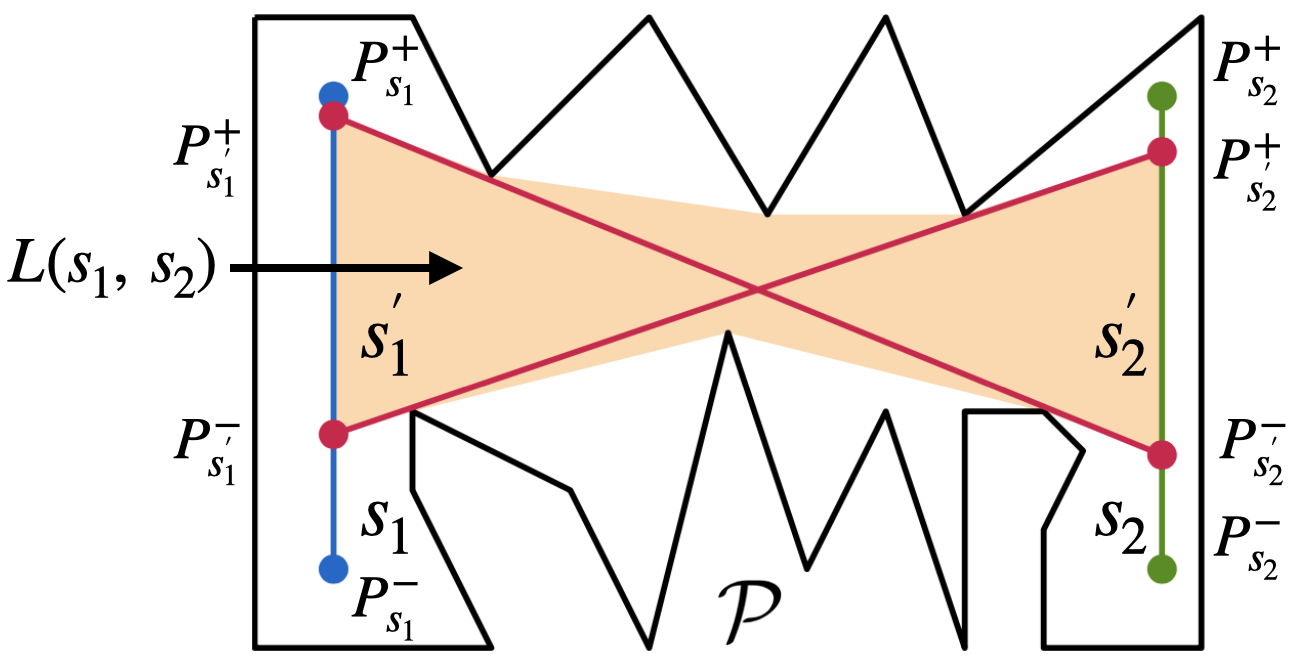}
        \caption{Visibility Glass of $s_1$ and $s_2$}
        \label{Fig1}
     \end{subfigure}
     \hfill
     \begin{subfigure}[b]{0.4\textwidth}
         \includegraphics[width=\textwidth]{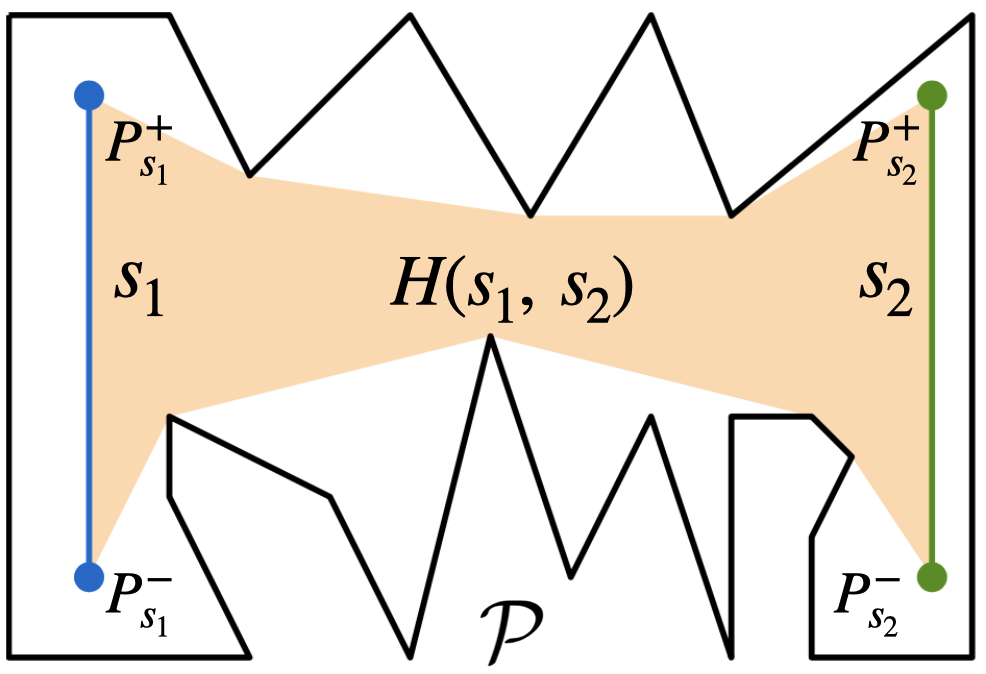}
         \caption{Hourglass of $s_1$ and $s_2$}
         \label{Fig2}
     \end{subfigure}
     \hfill
     \caption{\textbf{(a)} Visibility Glass~\cite{guibas1989optimal,eades2020trajectory} of two line-segments $s_1$ and $s_2$, $L(s_1, s_2)$, in $\mathcal{P}$, derived from $H(s_1, s_2)$. We denote the sub-line-segments of $s_1$ and $s_2$ contained in $L(s_1, s_2)$ as $s^{'}_{1}$ and $s^{'}_{2}$ respectively. Also, denote the first endpoint of $s_1$ as $P^{+}_{s_1}$ and the second one as $P^{-}_{s_2}$. Similarly, we define $P^{+}_{s_2}$ and $P^{-}_{s_2}$ for $s_2$, as well as $s^{'}_{1}$ and $s^{'}_{2}$. Note the bi-tangents specified as red line segments that specify $L(s_1, s_2)$ \textbf{(b)} Hourglass~\cite{guibas1989optimal,eades2020trajectory} of two line segments $s_1$ $\&$ $s_2$, called $H(s_1, s_2)$, in a simple polygon $\mathcal{P}$}
     \label{VGFig}
\end{figure}

Consider a simple polygon $\mathcal{P}$ with $n$ vertices. The maximal sub-polygon of $\mathcal{P}$  visible to a point $q \in \mathcal{P}$ is called the \textit{visibility polygon} of $q$, denoted by $\VP$. There are linear-time algorithms to compute $\VP$ when the viewer is a point \cite{lee1983visibility,joe1987corrections}. A line segment $l$ inside $\mathcal{P}$ is said to be completely visible from a line segment between $p, q \in \mathcal{P}$ if for every point $z \in l$ and for any point $w \in {pq}$, $w$ and $z$ are visible. 
A \textit{convex chain} consists of connected subsets of convex hulls due to Graham Convex Hull algorithm~\cite{graham1972efficient}. To construct a convex chain, we need to specify a starting and an ending point for the chain to run Graham's algorithm. Please refer to Appendix~\ref{VGAppendix} where we discuss Visibility Glass (\VG). Also, see Figure~\ref{VGFig} for further illustrations. Furthermore, the readers may refer to Appendix~\ref{sec:notations} as a summary of frequently used notations in this paper.  

\setlength{\textfloatsep}{10pt}
\setlength{\intextsep}{10pt}

\setlength{\textfloatsep}{10pt}
\setlength{\intextsep}{10pt}


\section{Completely Visible Trajectories}
\label{sec:completlyVisTrajs}

We will ultimately unfold different cases of our solution for $\TRVP$. Let us start by checking a special case: \textit{Can both trajectories entirely see one another?}

\begin{lemma}[$\CV$]
\label{lemma:completeVis}
    Let $m$ be the overall number of vertices in trajectories $\tau_q$ and $\tau_r$. Denote $n$ as the number of vertices of $\mathcal{P}$. There is an $\mathcal{O}(m \log n)$ time algorithm, requiring $\mathcal{O}(n)$ pre-processing time, that detects if every point on $\tau_q$ is visible to every point in $\tau_r$. 
\end{lemma}

\begin{proof}
     The algorithm uses the \textit{Graham algorithm}~\cite{graham1972efficient} to find the \textit{convex hull} of the set of all vertices (endpoints) in $\tau_q$ and $\tau_r$, $V_{q, r}$. Denote the set of vertices of the corresponding convex hull of $V_{q, r}$ as $C_{q,  r}$. Observe that a \underline{ray shooting query} \cite{chazelle1994ray} on the polygon $\mathcal{P}$ is required every time $C_{q,  r}$ gets a new vertex $v_c$. Note that performing such queries yield $\mathcal{O}(n)$ pre-processing time, as stated the Lemma above. If $\CV$ finds that the visibility on $v_c$ is blocked by the polygon $\mathcal{P}$, then \textit{complete visibility} between $\tau_q$ and $\tau_r$ no longer holds. However, if $\tau_q$ and $\tau_r$ see one another, the answer to the $\TRVP$ becomes rudimentary. Specifically, each entity can \textit{always} see the other entity, independent of the velocity. \qedsymbol{}
\end{proof}

\label{subsection::VG}
\setlength{\textfloatsep}{5pt}
\setlength{\intextsep}{5pt}

\section{Line-segment Trajectories with Given Velocities}
\label{sec:givenvelocities}

In this section, we reveal the details of our contribution to solving a simpler variant of $\TRVP$, as stated in Problem~\ref{TRGV}. We first provide a crucial part of our algorithm by reusing $\VG$. Second, we describe our visibility range tree idea to maintain the time intervals in $T_{q,r}$ (as defined in Problem~\ref{TRGV}). Note that in this section, we provide a solution that is not yet able to determine the exact boundaries of visibility for the moving entities. However, we resolve this in later sections.


\subsubsection{The $\VRT$ Algorithm}
\label{VTRAlg}

Observe that due to constraint (\ref{constraint1}), the trajectories are line segments. So, the overall number of trajectory vertices (endpoints) is exactly $4$, and thus constant. Therefore, referring to Lemma~\ref{lemma:completeVis}, it takes $\mathcal{O}(\log n)$ running and $\mathcal{O}(n)$ pre-processing time for $\VRT$ to check whether the complete visibility holds. If not, the following strategy can be taken: Recalling from Subsection~\ref{subsection::VG}, the visibility glass ($\VG$) of two line segments $s_1$ and $s_2$ in a simple polygon $\mathcal{P}$ were discussed. Note that currently the corresponding trajectories of $q$ and $r$, ${\tau}_q$ and ${\tau}_r$, are assumed to be line-segments. Accordingly, using $\VG$ technique would narrow down the problem to the sub-segments of ${\tau}^{*}_{q} \subset {\tau}_q$ and ${\tau}^{*}_{r} \subset {\tau}_r$ which $\VG$ technique specifies (see Subsection~\ref{subsection::VG}). So, if there are two moving entities $q$ and $r$ with constant velocities $v_q(t) = C_0$ and $v_r(t) = C_1$ on ${\tau}_r$ and ${\tau}_q$, $\VRT$ aims only for the parts that it is possible for the entities to see one another. In particular, the areas outside of $L({\tau}_q, {\tau}_r)$, the $\VG$ of ${\tau}_q$ and ${\tau}_r$, can be discarded. Let the number of vertices in $\mathcal{P}$ be $n$. Therefore, it takes $\mathcal{O}(n)$ running time to find $L({\tau}_q, {\tau}_r)$. Note that if $L({\tau}_q, {\tau}_r)$ becomes empty, the algorithm can stop.

Consider the integrality assumption (as appeared in constraint (\ref{constraint4})). Given two trajectories $\tau_r$ and $\tau_q$ as segments in $\mathcal{P}$, $\VRT$ first obtains $L(\tau_q, \tau_r)$ in $\mathcal{O}(n)$ time. The algorithm would then divide each sub-trajectory resulted from calculating $L(\tau_q, \tau_r)$ into \textit{small} sub-segments like $\tau^{*}_{q_k}$ and $\tau^{*}_{r_w}$ s.t. the length of each \textit{small} segment is $d$, $k \in [1, \ \frac{{len}(\tau^{*}_{q})}{s}] \ \& \ w \in [1, \  \frac{{len}(\tau^{*}_{r})}{d} ]$, and ${len}(s)$ is the length of an arbitrary segment $s$. For convenience, $\VRT$ can rotate $\mathcal{P}$ until $\tau^{*}_{q}$ becomes parallel to the x-axis. Therefore, all points on $\tau^{*}_{q}$ have the same y-coordinate and $\VRT$ can \textit{conceptually} consider some of its computations in 1D space. More specifically, the algorithm builds a 1D Range Tree $\mathcal{T}_q$ based on the \textit{small} segments created before on $\tau^{*}_{q}$. Each node $\nu \in \mathcal{T}_q$ stores the following information: The smallest and the largest x-coordinates covered by $\nu$ as $R(\nu) = [x_s, x_l]$, the left and the right most segments included in $\nu$, and an empty list $T_{\nu}$. As mentioned, the algorithm divides $\tau^{*}_{q}$ and $\tau^{*}_{r}$ into segments like $\tau^{*}_{q_k}$ and $\tau^{*}_{r_w}$. Accordingly, it moves $r$ (similarly $q$) along every $\tau^{*}_{r_w}$ for $w \in [1, \  \frac{{len}(\tau^{*}_{r})}{d} ]$ and runs queries on $\mathcal{T}_q$ to find every $\tau^{*}_{q_k}$ visible to the \textit{current} $\tau^{*}_{r_w}$. Each time $\VRT$ executes a query on $\mathcal{T}_q$, it specifically uses the following procedure: Starting from the root, first, check if the current node represents a range completely visible to $r$. If yes, add the current \textit{timestamp} (we will define this rigorously later in this section) to $T_{\nu}$ and return. If both the segments in $\nu$ are invisible to $q$, then do nothing and return. Otherwise, invoke the same procedure for both children of $\nu$. It can be proved by induction that there are at most two nodes per level of $\mathcal{T}_q$ that are completely visible to $q$, as shown in~\cite{qiao2016range,zhang2022approximate}. Recall that every node $\nu \in \mathcal{T}_q$ stores the left and the right most segments included in $\nu$. So, it takes a constant number of \textit{ray shooting operations} to see if each of the two segments is completely visible to $r$. In particular, $\VRT$ aims to find the first intersection point of a ray with $\mathcal{P}$ s.t. the ray crosses the endpoints of both segments stored in $\nu$.
    
Observe that there can be cases in which a segment in $\nu$ is \textit{partially} visible to $q$. So, the challenge mentioned for the straightforward solution, regarding having an infinite number of points between two endpoints of a segment, appears again. To resolve it, let $\varepsilon$ be a given tolerable error rate. To estimate the visible area on a segment, $\VRT$ uses the binary search idea as follows: Consider the endpoints of the segment as $e_s$ and $e_e$. Pick the middle point between $e_s$ and $e_e$ and call it $e_m$. W.l.o.g suppose that $e_e$ is not visible, but $e_s$ is. Then, if $e_m$ is also not visible, set $e_e = e_m$ and calculate a new $e_m$. Similarly, if $e_m$ is visible, change $e_s = e_m$ and calculate the new $e_m$. The procedure stops only if the euclidean distance between $e_s$ and $e_e$ is less than or equal to $\varepsilon$.
Therefore, $e_m$ can be returned as an estimate within a factor of $\varepsilon$ of the exact intersection point. To extract the time intervals in which $\tau^{*}_{r}$ becomes visible to $q$ (similarly, $\tau^{*}_{q}$ for $r$), $\VRT$ can pick a strategy as follows: As $v_q(t) = C_0$ and $v_r(t) = C_1$, it is easy to define the \textit{timestamp} for each moving entity. Let $t_q = \frac{1}{L C_0}$ and $t_r = \frac{1}{L C_1}$ be the corresponding timestamps of $q$ and $r$, respectively. So, each time $\VRT$ inserts a \textit{timestamp} in $T_{\nu}$ s.t. $\nu$ is a range tree node, it would be the total time for a moving entity to reach the current position considered for it. So, for a given range on a segment, the algorithm can determine the time intervals it became visible.

\begin{lemma}
\label{lemma:VRT}
    The Visibility Range Tree Algrithm($\VRT$) solves Problem~\ref{TRGV} with the error rate of $\frac{1}{L} \leq \varepsilon$, offering $\mathcal{O}(t \log t \log n)$ running time, where $t = \max \{ {len}(\tau^{*}_{q}) \cdot L, \ {len}(\tau^{*}_{r}) \cdot L \}$. Given a query small segment with length $\frac{1}{L}$ and/or a \textit{range} (i.e. a set of contiguous small segments), the algorithm offers $\mathcal{O}(\log t  + k)$ time to report all timestamps that the query segment or range becomes visible s.t. $k$ is the number of such timestamps. The space consumption of the algorithm is $\mathcal{O}(t \log t)$ and it requires $\mathcal{O}(n)$ pre-processing time (see Lemma~\ref{lemma:completeVis}).
\end{lemma}

\begin{proof}
   Please refer to Appendix~\ref{proof:lemmaVRT}.
\end{proof}

\section{Trajectory Visibility on a Simple Polygon}
\label{sec:arbitraryvelocity}

In this section, we discuss the details of our contribution to solving $\TRVP$. Initially, constraints~(\ref{constraint4}, \ref{constraint5}) will be removed to present a simpler version of the final algorithm. For this purpose, we reuse some techniques discussed in Lemma~\ref{lemma:VRT}, and also add more details to improve the results. We then discuss the final version by nullifying constraint~(\ref{constraint1}). 

\subsection{Line-segment Trajectories}
\label{subsec:FirstVariant}
Note that  constraints (\ref{constraint4}, \ref{constraint5}) no longer hold. However, constraint (\ref{constraint1}) is to be removed later. Accordingly, we construct $L(\tau_q, \tau_r)$ inside $\mathcal{P}$ in $\mathcal{O}(n)$ time (similar to Lemma~\ref{lemma:VRT}). Observe that if $L(\tau_q, \tau_r)$ becomes empty, the algorithm stops. Otherwise, it is crucial to note that regardless of the possible values for $v_q(t)$ and $v_r(t)$ (see Problem~\ref{TRSP}), the visible region on each trajectory remains the same. So, $q$ always sees the same sub-trajectory(s) of $\tau^{*}_{r}$, independent from the value of $v_q(t)$ and vise versa. Therefore, such pairs of visible sub-trajectories on $\tau^{*}_{q}$ and $\tau^{*}_{r}$ should be specified. Then, \textit{ranges} of velocities for $q$ and $r$ have to be determined so that they fall into such areas visible to one another.

\begin{definition}[Mapping Positions to Velocities]
    \label{timestampVelocity}
    If two sub-trajectories $\tau^{*}_{q}$ and $\tau^{*}_{r}$ are visible to each other, then \textit{ranges} like $[C_{i}, C_{j}]$ and $[C_{k}, C_{t}]$ can be found, in which $j > i$ and $t > k$, for $v_q(t)$ and $v_r(t)$ respectively. These ranges are computed in a way that if a \textit{unit of time} (mostly a second, if velocities are in terms of $m/s$) passes, the entity moves forward on its corresponding trajectory as large as its velocity.
\end{definition}

For example, if $v_q(t) = C_i$, $q$ would stand on the beginning of $\tau^{*}_{q}$ (say $x$), and if $v_r(t) = C_t$, then $r$ would be at the end of $\tau^{*}_{r}$ (say $y$). So, $q$ and $r$ can see each other from points $x$ and $y$. Since constraint (\ref{constraint4}) no longer holds, we cannot use the same approach as Section~\ref{sec:givenvelocities} and define $\frac{1}{L} \leq d$. Accordingly, we use a different approach using the properties of $L(\tau_q, \tau_r)$. 

\subsubsection{The $\LRTV$ Algorithm}
\label{LRTVAlg}
The $\LRTV$ algorithm solves the problem in Definition~\ref{TRSP}, restricted to constraints (\ref{constraint1}, \ref{constraint2}, \ref{constraint3}), and based on Definition~\ref{timestampVelocity}. It first requires checking whether complete visibility holds as appeared in Lemma~\ref{lemma:completeVis}. Otherwise, referring to Figure~\ref{Fig1}, the algorithm considers the shortest path between $P^{+}_{s^{'}_{1}}$ and $P^{+}_{s^{'}_{2}}$ (likewise, the shortest path between  $P^{-}_{s^{'}_{1}}$ and $P^{-}_{s^{'}_{2}}$), and denotes it as $S$. It is easy to observe that the first reflex vertex blocking the sight of $P^{+}_{s^{'}_{1}}$ on $s^{'}_{2}$ (if exists) is on the bi-tangent, drawn from $P^{+}_{s^{'}_{1}}$ to $P^{-}_{s^{'}_{2}}$. Denote the aforementioned bi-tangent as $b_1$ and the other as $b_2$.
Therefore, traversing on $S$, one can \textit{extend} the edge, say $e$, between two consecutive vertices until $e$ intersects $b_1$ and/or $b_2$. Observe that each intersection with a bi-tangent, for instance $b_1$, forms a wedge $\Lambda_{e, b_1}$ s.t. $\Lambda_{e, b_1}$ intersects at two points with $\tau^{*}_{q}$, say $x_1$ and $x_2$, and at two other points with $\tau^{*}_{r}$, say $x_3$ and $x_4$. W.l.o.g. suppose that $x_2 > x_1$ and $x_4 > x_3$. 
It is clear that $\overline{x_1 x_2}$ and $\overline{x_3 x_4}$ are completely visible to each other. Recall $\tau^{*}_{q}$ and $\tau^{*}_{r}$ are denoted as sub-trajectories of $\tau_q$ and $\tau_r$. In this case, $\LRTV$ considers $\overline{x_1 x_2}$ and $\overline{x_3 x_4}$ as $\tau^{*}_{q}$ and $\tau^{*}_{r}$, respectively. Thus, it is easy to find $[C_{i}, C_{j}]$ and $[C_{k}, C_{t}]$.
However, there are two challenges facing this approach. First, in the worst case, there can be $\mathcal{O}(n)$ wedges like $\Lambda_{e, b_1}$. Second, the \textit{ranges}\footnote{In fact, one can denote $[x_3, x_4]$ and $[x_1, x_2]$ as ranges on the trajectories} such wedges specify on $\tau^{*}_{q}$ and $\tau^{*}_{r}$ might heavily overlap with one another. To resolve this, the \textit{endpoint tree}~\cite{qiao2016range} can be used as follows: Once finding all wedges, it is easy to find the velocities that correspond to their ends, similar to what was mentioned earlier in the example regarding $[C_{i}, C_{j}]$ and $[C_{k}, C_{t}]$. The algorithm can then build two endpoint trees $\mathcal{T}_q$ and $\mathcal{T}_r$ based on the velocities calculated in the previous step. Observe that a range like $[x_1, x_2]$ on $\tau^{*}_{q}$ \textit{sees} $[x_3, x_4]$ on $\tau^{*}_{r}$, and $\LRTV$ maps $[x_1, x_2]$ to $[C_{i}, C_{j}]$ and $[x_3, x_4]$ to $[C_{k}, C_{t}]$. It is clear that ranges $[x_1, x_2] \in \tau^{*}_{q}$ and $[x_3, x_4] \in \tau^{*}_{r}$ are visible to each other. So, the following strategy can be used: For each range on $\tau^{*}_{q}$ (similarly on $\tau^{*}_{r}$), find its corresponding \textit{participants} on $\mathcal{T}_r$ via the same way discussed in~\cite{qiao2016range}. $\LRTV$ then determines the visible area of every participant using the same reflex vertices that specify the range.
So, the algorithm can complete constructing $\mathcal{T}_q$ and $\mathcal{T}_r$. Accordingly, $\mathcal{T}_q$ allows answering the following queries(see Problem~\ref{TRSP} and Definition~\ref{timestampVelocity}): $(i)$ \textit{What is the range of velocity that viewers become visible if they want to meet within given ranges (in terms of position)?} ($ii$) \textit{What are the sub-trajectories that $q$ and $r$ become visible to each other if they have given velocities $C_0$ and $C_1$?}

\begin{theorem}
\label{theorem:lineSegProb2}
The $\LRTV$ algorithm provides $\mathcal{O}(n \log n)$ running time, solving the problem presented by Definition~\ref{TRSP}, restricted to constraints (\ref{constraint1}, \ref{constraint2}, \ref{constraint3}), and based on Definition~\ref{timestampVelocity}. The algorithm offers $\mathcal{O}(\log n)$ query time, $\mathcal{O}(n)$ pre-processing time (please refer to Lemma \ref{lemma:completeVis}), and requires the space complexity of $\mathcal{O}(n \log n)$.
\end{theorem}

\begin{proof} 
    The correctness and the complexity analysis of the $\LRTV$ (see Sub-subsection~\ref{LRTVAlg}) are as follows:

     \textbf{Correctness.} W.lo.g consider $\nu \in \tau_q$ a point (a trajectory vertex or any other point) visible for at least one point like $x$ on $\tau_r$. Denote the data structure $\LRTV$ finally generates as $L$. Let us assume that $\nu$ will not be reported for $x$ if we query $L$ with velocity $v_r$, which brings the moving entity $r$ to $x$, according to Definition~\ref{timestampVelocity}. Notice that if $\nu$ is not visible for at least one point in $\tau_r$, it will not appear in the $\VG$ that $\LRTV$ constructs. On the other hand, note that only the reflex vertices block the visibility on $\nu$. Note that $\LRTV$ traverses on $S$ and \textit{extends} the edge, say $e$, between two consecutive vertices until $e$ intersects $b_1$ and/or $b_2$. So, the \textit{wedge} in which $\nu$ falls will eventually be considered by $\LRTV$. Accordingly, a point like $\nu$ does not exist.
    
    \textbf{Time and Space Complexity.} $\LRTV$ builds an endpoint tree using the velocities it calculates. Observe that there can be $\mathcal{O}(n)$ ranges that the $\LRTV$ creates. Accordingly, the construction time and memory consumption of the endpoint tree becomes bounded by $\mathcal{O}(n \log n)$. Also, since the participants on the endpoint tree are bounded by $\mathcal{O}(\log n)$, $\LRTV$ thus preserves the overall running time of $\mathcal{O}(n \log n)$. Note that the query time of the endpoint tree is bounded by $\mathcal{O}(\log n)$.
\end{proof} 
 
\subsection{Restricted Non-constant Complexity Trajectories}
\label{nonSegmentRestricted}

Recall from \ref{subsec:FirstVariant} that we require constructing $L(\tau_q, \tau_r)$ inside $\mathcal{P}$, and if $L(\tau_q, \tau_r)$ becomes empty, the algorithm stops. Note that if we keep constraint (\ref{constraint1}), the \textit{endpoints} of $\tau_q$ and $\tau_r$ suffice to form $L(\tau_q, \tau_r)$ properly. However, removing constraint (\ref{constraint1}) yields \textit{arbitrary} forms of trajectories that \textit{endpoints} would no longer specify \textit{boundaries} of the trajectories. That is, the endpoints can no longer provide an insight into the area of $\mathcal{P}$ in which a trajectory exists. Thus, computing the union of the shortest paths between all pairs of points in $\tau_q$ and $\tau_r$ requires considering all of their line segments. Therefore, one cannot directly use the method discussed in Subsection~\ref{subsection::VG} to construct $L(\tau_q, \tau_r)$.

To resolve this, first, observe that there can be reflex vertices in $\mathcal{P}$ as the only elements that block the visibility between the trajectories. Second, since we still keep constraint (\ref{constraint2}), we can observe a crucial item regarding visibility. Consider two moving entities $q$ and $r$ and their corresponding trajectories $\tau_q$ and $\tau_r$ respectively. Suppose that $q$ is moving from point $p_i \in \tau^{*}_{q}$ to $p_j \in \tau^{*}_{q}$ s.t. $\tau^{*}_{q} \subset \tau_{q}$, $\tau^{*}_{r} \subset \tau_{r}$, and $\tau^{*}_{q}$ is visible from $\tau^{*}_{r}$. Then, there cannot be a point like $p_k \in \tau^{*}_{q}$ and between $p_i$ and $p_j$ s.t. $p_k$ is invisible from $\tau^{*}_{q}$. So, the visibility remains \textit{continuous}, as long as it exists.

\begin{remark}
\label{remark:restrictedNonSegRemark}
We would later introduce the \textit{connecting vertices} in this Subsection. Note that we currently restrict the trajectories $\tau_q$ and $\tau_r$ to the following: $(*)$ There is at least one vertex $x \in \tau_q$ or $x \in \tau_r$ that has \textit{no visibility} on all points on the other trajectory $(**)$ We can find connecting vertices $v^{1}_{q}, v^{2}_{q} \in \tau_q$ and $v^{1}_{r}, v^{2}_{r} \in \tau_r$ s.t. constructing $L(\overline{v^{1}_{q} v^{2}_{q}}, \overline{v^{1}_{r} v^{2}_{r}})$ helps solving $\TRVP$, as if calculating $\VG$ for all pairs of line-segments in both trajectories.
\end{remark}

According to the above observations, we use a scan denoted as $G_s$, similar to the procedure in \textit{Graham algorithm}~\cite{graham1972efficient,VAEZI201922} on the set of vertices of $\tau_q$ and $\tau_r$. Denote this set as $V_{q, r}$. To start the scan, initially find a trajectory vertex that does not have visibility on the vertices of the other trajectory. Specifically, when starting $G_s$ from a trajectory vertex $\nu_q \in \tau_q$ and then picking the next vertex as $\nu_r \in \tau_r$ (or vice versa), while $\nu_q$ sees $\nu_r$, choose another starting point for the scan on the opposite \textit{direction} compared to $G_s$. More specifically, the starting point can be determined as follows:

\begin{lemma}
\label{lemma:findInvPoint}
    There is an algorithm, called $\SPI$, that finds a starting point in $\mathcal{O}(n + m)$ time, and reports if both trajectories are completely invisible to each other. In more formal words, $\SPI$ finds out if no pairs of points in $\tau_q$ and $\tau_r$ can see each other.
\end{lemma}

\begin{proof}
    $\SPI$ picks a vertex $v \in \tau_q$ (or $\tau_r$) and computes a $\VP$ (see Subsection~\ref{subsec::Defs}) from $v$ in $\mathcal{O}(n)$ time~\cite{joe1987corrections,lee1983visibility}. The algorithm also considers the vertices of $\tau_q$ and $\tau_r$ while constructing $\VP$. However, once $\SPI$ detects a trajectory vertex or intersects a trajectory edge for adding to $\VP$, the algorithm marks it and pretends that it did not meet such a vertex. $\SPI$ then continues with the unmarked vertices until it forms a $\VP$ s.t. it no longer meets a trajectory vertex and intersects a trajectory edge no more while constructing the $\VP$. Observe that the area inside each $\VP$ is disjoint with all other $\VP$s the algorithm creates. Thus, $\SPI$ can find the starting point in linear time in terms of the number of trajectories and polygon vertices. The algorithm can also detect the case that both trajectories are \textit{completely invisible} for each other. Observe that if all $\VP$s detect the vertices (also intersections with edges) of only one trajectory, then complete invisibility holds. 
\end{proof}

\noindent Following the direction of $G_s$ from the picked starting point, we require finding two crucial edges that connect the trajectories. Let us call the latter edges $e_{1}$ and $e_{2}$. So, the scan procedure $G_s$ will stop as soon as it finds $e_{1}$ and $e_{2}$. To avoid visiting vertices more than once, remove each vertex from $V_{q, r}$ when visiting it.

\begin{definition}[Connecting Vertices]
    \label{definition:ConnectingVertices}
    Let $e_{1}$ and $e_{2}$ be the edges that connect the trajectories when following the direction of $G_s$ from the picked starting point. Denote the vertices on $e_{1}$ and $e_{2}$ as $v^{1}_q$, $v^{2}_q$, $v^{1}_r$, and $v^{2}_r$. More specifically, $v^{1}_q$ and $v^{1}_r$ are the endpoints of $e_{1}$. Also, $v^{2}_q$ and $v^{2}_r$ are the endpoints of $e_{2}$. Let us call these vertices as \textit{connecting vertices}.
\end{definition}

Observe that there can be edges like $e_{\mathcal{P}}$ or reflex vertices like $r_{\mathcal{P}}$ of $\mathcal{P}$ that might block $e_{1}$ and/or $e_{2}$. Therefore, once picking a \textit{connecting vertex} like $v_c$, using the $G_s$, run a \textit{ray shooting} query on $\mathcal{P}$ in $\mathcal{O}(\log n)$ time to find out whether $v_c$ sees a connecting vertex on the other trajectory. Accordingly, when the visibility for $v_c$ is not blocked, the connecting vertices can be labeled as follows: 

\begin{definition}[Upper and Lower Points]
\label{definition:upperAndLower}
    Considering Definition~\ref{definition:ConnectingVertices}, if there is no edge or reflex vertex blocking the sight of $v^{1}_q$ on $v^{1}_r$ (as well as $v^{2}_q$ and $v^{2}_r$), pick $v^{1}_q$ and $v^{1}_r$ and call them \textit{upper points} on $\tau_q$ and $\tau_r$ respectively (Naturally, we refer to $v^{2}_q$ and $v^{2}_r$ as \textit{lower points}). Note that the word \textit{upper} (and \textit{lower}) \underline{is a naming convention} and does \textit{not} indicate a particular direction.
\end{definition}

On the other hand, if $e_{\mathcal{P}}$ or $r_{\mathcal{P}}$ exists, change the strategy as follows: After eventually finding the reflex vertex $r_{\mathcal{P}}$, add $r_{\mathcal{P}}$ to $V_{q, r}$ and start $G_s$ again from $r_{\mathcal{P}}$, considering the same \textit{direction} $G_s$ had.
There can be two cases then when $G_s$ resumes scanning: First, if the next vertex of $r_{\mathcal{P}}$ ($v_{next}$), due to $G_s$, belongs to the other trajectory and is visible to $r_{\mathcal{P}}$, pick the most recently met vertex and $v_{next}$ as connecting points. Then continue the scan from $v_{next}$. However, hitting another vertex on the same trajectory yields the second case. So, the $G_s$ continues from $v_{next}$.
Therefore, if the connecting vertices are found, the hourglass and $\VG$ can be constructed. As a result, the connecting vertices are \textit{representative} points of the trajectories, for which constructing $L(\overline{v^{1}_q v^{2}_q}, \overline{v^{1}_r, v^{2}_r})$ is equivalent to constructing a $\VG$ for all pairs of sub-trajectories in $\tau_q$ and $\tau_r$. 

\subsubsection{The $\RTRV$ Algorithm}
\label{RTVAlg}

$\RTRV$ solves Problem~\ref{TRSP}, subject to the constraints (\ref{constraint2},  \ref{constraint3}), and Remark~\ref{remark:restrictedNonSegRemark}. Firstly, $\RTRV$ checks whether the complete visibility holds between the trajectories (according to Lemma~\ref{lemma:completeVis}). If not, it takes the following strategy: Using Lemma~\ref{lemma:findInvPoint}, find a starting point or report that the trajectories are not visible to each other. If the starting point exists, find the upper and lower points, as defined in Definition~\ref{definition:upperAndLower}. Now consider the shortest paths $S_q$ and $S_r$ between the upper and the lower points found on $\tau_q$ and $\tau_r$, respectively. If $S_q$ and $S_r$ are line-segments, use the bi-tangents $b_1$ and $b_2$ on $L(\overline{v^{1}_q v^{2}_q}, \overline{v^{1}_r, v^{2}_r})$, and \textit{extend} them until they hit $\mathcal{P}$. $\RTRV$ thus creates a set of intersections of the bi-tangents with the trajectories and all trajectory vertices that fall within the visible area that the bi-tangents specify on $\mathcal{P}$. Denote the intersection points and visible vertices on $\tau_q$ and $\tau_r$ as ${vis}(\tau_q)$ and ${vis}(\tau_r)$, respectively. To obtain the velocity required to reach a point like $\nu_i$ in ${vis}(\tau_q)$ or ${vis}(\tau_r)$ within a timestamp, $\RTRV$ calculates the distance $d(\nu_i)$. Specifically, $d(\nu_i)$ is the distance one should travel from the beginning of $\tau_q$ or $\tau_r$ to reach $\nu_i$. Suppose $\nu_i \in {vis}(\tau_q)$ (note that the algorithm uses the same approach if $\nu_i \in {vis}(\tau_r)$). The algorithm then checks which one of the lines that cross the segments $\overline{\nu_i P^{+}_{\tau_r}}$ and $\overline{\nu_i P^{-}_{\tau_r}}$ intersects $\overline{P^{+}_{\tau_q} P^{-}_{\tau_q}}$(\textit{case 1}). Observe that $\RTRV$ checks the intersection with $\overline{P^{+}_{\tau_q} P^{-}_{\tau_q}}$ to see if $\mathcal{P}$ blocks the visibility on lines that cross $\overline{\nu_i P^{+}_{\tau_r}}$ and $\overline{\nu_i P^{-}_{\tau_r}}$. In particular, checking if $\overline{\nu_i P^{+}_{\tau_r}}$ and $\overline{\nu_i P^{-}_{\tau_r}}$ fall within the $L(\overline{v^{1}_q v^{2}_q}, \overline{v^{1}_r, v^{2}_r})$. On the other hand, if both lines intersect $\overline{P^{+}_{\tau_q} P^{-}_{\tau_q}}$, then $\RTRV$ can consider both (\textit{case 2}). Lastly, the algorithm can discard $\nu_i$ when unable to find such lines from $\nu_i$ (\textit{case 3}). However, crucial items remain: 

        For intersection points like $\nu_i$ between a bi-tangent and $\tau_q$, $\RTRV$ performs ray shooting queries on the bi-tangents and discards $\nu_i$ if a point on $\mathcal{P}$ blocks its visibility. The latter intersections will then be marked and used later. Moreover, note that reflex vertices like $r$ can block the visibility to trajectory vertices in ${vis}(\tau_q)$. However, there is no need to use a ray-shooting query when $r$ exists. The algorithm can first insert all trajectory vertices between $b_1$ and $b_2$ in a set called $\mathcal{M}$. Note that $\mathcal{M}$ remains for the later steps. Second, $\RTRV$ traces the vertices of $\mathcal{P}$ from the point that the extension of $b_1$ (or $b_2$) hits $\mathcal{P}$. The algorithm then checks \textit{case 1,2,3} from $r$, considering the line $l_r$ that crosses $r$ and $P^{+}_{\tau_q}$ or $P^{-}_{\tau_q}$. If two such lines were found, $\RTRV$ picks the one with the tangent closer to the tangent of $b_1$ (from which we started tracing). Once found $r$, the algorithm keeps tracing reflex vertices in $\mathcal{P}$ that fall between $b_1$ and $b_2$ and their intersections with $\mathcal{P}$. If $\RTRV$ finds a reflex vertex $r^{'}$ for which \textit{case 3} holds, $r^{'}$ gets discarded. Also, if a reflex vertex falls out of the area that $l_r$ specifies, it again gets discarded. Observe that if $l_r$ intersects $\mathcal{P}$ only at $r$, it cannot block the visibility of trajectory and polygon vertices. Verifying this only needs checking if $l_r$ falls between the two edges that intersect $r$. Also, if $l_r$ intersects $\mathcal{P}$ outside of the area between the bi-tangents, $\RTRV$ can again ignore the polygon and trajectory vertices that are not between $b_1$ and $b_2$. Otherwise, the algorithm computes $l_{r^{'}}$ similar to $l_r$ and then focuses on the trajectory vertices that fall between $l_r$ and $l_{r^{'}}$. Note that if $\RTRV$ does not find $r^{'}$ and/or $r$, it instead uses $b_2$ and/or $b_1$, respectively. Moreover, it considers $l_{r^{''}}$ instead of $r$, if there is a reflex vertex $r^{''}$ which was found before $r$, while $r$ does not block the visibility or if $r$ is not found.

        On the other hand, it takes constant time to check if trajectory vertices like $\mu \in \mathcal{M}$ and polygon vertices like $\nu_i$, fall not in the area between $l_r$ and the corresponding edge of $r$ that leaves it. $\RTRV$ can similarly check for $l_{r^{'}}$ and $e_{r^{'}}$ s.t. $e_{r^{'}}$ is the edge that leaves $r^{'}$. When the algorithm finishes checking $\mu$, it removes $\mu$ from $\mathcal{M}$ in order to avoid meeting $\mu$ more than once. In addition, recall that tracing $\mathcal{P}$ was started from the intersection of $b_1$ with $\mathcal{P}$. Accordingly, $\RTRV$ can detect which edges leave $r$ and $r^{'}$ and fall between $l_r$ and $l_{r^{'}}$. So, the algorithm continues and finds the intersection points of trajectory edges (if exist) with $l_r$ and/or $l_{r^{'}}$ and checks cases 1, 2, and 3 for such points. $\RTRV$ marks such points different from the intersections of bi-tangents with the trajectories and uses them later. Finally, the algorithm stops when it finds all visible trajectory vertices and the intersections of all lines like $l_r$ with trajectory vertices or edges.

        Observe that, to find all the velocities that moving entity $q$ can have, when a query velocity $v_r$ is given for the entity $r$, $\RTRV$ needs to have a different approach compared to Subsection~\ref{subsec:FirstVariant}. The algorithm first needs to determine if $v_r$ corresponds to a point $t \in \tau_r$ s.t. $t$ is inside an area where a reflex vertex does not block its visibility. The algorithm has already found all intersection points and trajectory vertices on $\tau_r$ whose visibility is not blocked. Also, they can form sub-trajectories of $\tau_r$. Accordingly, $\RTRV$ specifies all such sub-trajectories and marks their corresponding vertices and intersection points. Recall that the algorithm marked the intersections of the bi-tangents with the trajectories different than the intersections of all lines like $l_r$ with the trajectories. So, if $\RTRV$ traces the found trajectory vertices and the intersection points, it can use the marks and detect the sub-trajectories. 

        If the algorithm detects a trajectory vertex that falls on the same line with an intersection point and vice versa, it still has not reached the end of a sub-trajectory. So it marks the detected vertex or intersection as a vertex of the same sub-trajectory. Also, if two intersection points with different marks are detected, $\RTRV$ would remain on the same sub-trajectory and use the same mark. Otherwise, the algorithm can mark the vertex or intersection point it reached as the vertex of a new sub-trajectory. It can then sort all vertices and intersection points based on their distance from the beginning of the trajectory.
        Observe that if the predecessor and the successor of $v_r$ are not on the same sub-trajectory, then $v_r$ does not correspond to a point with sight on the other trajectory. If not, $\RTRV$ can search among the trajectory vertices and intersections on $\tau_q$ and detect the visible ones.

        Note that as we discussed for the bi-tangents, $\overline{t P^{+}_{\tau_q}}$ and/or $\overline{t P^{-}_{\tau_q}}$ can intersect non-constant number of trajectory segments on $\tau_q$. 
        To resolve this, use half-plane queries based on the half-planes that cross $\overline{t P^{+}_{\tau_q}}$ and/or $\overline{t P^{-}_{\tau_q}}$, to detect all intersections and trajectory vertices that fall within the visible area for the moving entity $r$. Denote the number of all such intersections and vertices as $k$.
        Note that if both $\overline{t P^{+}_{\tau_q}}$ and $\overline{t P^{-}_{\tau_q}}$ exist, the algorithm should calculate their intersection. 
        The half-plane queries would then specify $k$ and $k^{'}$ vertices and intersections. Thus, the intersection can be found. 
        In addition, it is rudimentary to compute the velocity ranges according to the results obtained: Based on $\overline{t P^{+}_{\tau_q}}$ and/or $\overline{t P^{-}_{\tau_q}}$, find that if the corresponding segment of each vertex or intersection point falls within the visible area. If not, specify the intersection of each segment with $\overline{t P^{+}_{\tau_q}}$ and/or $\overline{t P^{-}_{\tau_q}}$.

        On the other hand, if $S_q$ and $S_r$ are \textit{not} line-segments, $\RTRV$ first computes the $H(S_q, S_r)$. Recall from Subsection~\ref{subsection::VG} that the upper chain is from $v^{1}_{q}$ to $v^{1}_{r}$, and the lower chain is from $v^{2}_{q}$ to $v^{2}_{r}$. For a connecting vertex $x$, denote its adjacent reflex vertex on the upper or lower chain as $r_x$. Note that each vertex like $x$ can have at most one such reflex vertex adjacent to it, which is also not equal to the corresponding reflex vertex of the other connecting vertices. Moreover, $x$ can have another adjacent reflex vertex $r^{'}_{x}$ on $S_q$ or $S_r$. One can extend $\overline{x r^{'}_{x}}$ and $\overline{y r^{'}_{y}}$ until they intersect in a point like $i_{x, y}$, s.t. $y$ is the other connecting vertex on the same trajectory with $x$. It is easy to use a ray shooting query to check if $i_{x, y}$ is in $\mathcal{P}$ or not. If $i_{x, y} \in  \mathcal{P}$, the algorithm considers extending $\overline{i_{x, y} r_x}$ and $\overline{i_{x, y} r^{'}_{x}}$ until they hit $\mathcal{P}$. Note that this is again doable, using a ray shooting query on each segment. If $i_{x, y}$ is not in $\mathcal{P}$, $\RTRV$ can only consider the extensions from their intersections with $\mathcal{P}$. Lastly, compute the extension of $\overline{x r_x}$ and $\overline{x r^{'}_{x}}$ until they hit $\mathcal{P}$. Then compute all trajectory vertices falling between the extensions mentioned, using the same approach in Lemma~\ref{lemma:findInvPoint} and continue like the previous case.

\begin{theorem}
    \label{theorem:RestrictedTRSP}
    The $\RTRV$ algorithm provides $\mathcal{O}(n + m (\log m + \log n))$ running time to solve Problem~\ref{TRSP}, subject to the constraints (\ref{constraint2},  \ref{constraint3}), and Remark~\ref{remark:restrictedNonSegRemark}. The algorithm requires $\mathcal{O}(n)$ pre-processing time and only $\mathcal{O}(n + m)$ space. The query time is bounded by $\mathcal{O}(\log m + k)$.
\end{theorem}

\begin{proof}
    Please refer to Appendix~\ref{proof:theoremRestrictedTRSP}.
\end{proof}

\subsection{Non-constant Complexity Trajectories}
\label{subsection:NonSegTrajs}

Recall from Remark~\ref{remark:restrictedNonSegRemark} that we assumed there is at least one vertex on a trajectory that does not have visibility on all points of the other trajectory. And, there are exactly four connecting vertices, as discussed. At this stage, we are ready to lift the previous assumptions.

\subsubsection{The $\PNST$ Algorithm}
\label{PNSTAlg}

$\PNST$ solves Problem~\ref{TRSP}, subject to the constraints (\ref{constraint2}, \ref{constraint3}). The algorithm divides its strategy into two phases: $(1)$ Finding the upper points and $(2)$ Determining the rest of the connecting points based on the found upper points. To start phase $(1)$, it reuses the way discussed for $\RTRV$, but with the following modifications: If all vertices on $\tau_q$ see at least one point from $\tau_r$, start $G_s$ on $\tau_q$, and pick the first vertex $G_s$ chooses as an upper point. Then continue $G_s$ like before while not hitting a reflex vertex $r_{\mathcal{P}}$ (or indirectly, its corresponding edge) on $\mathcal{P}$. Yet, if $\PNST$ hits $r_{\mathcal{P}}$ and $r_{\mathcal{P}}$ has visibility, it picks the next vertex of $r_{\mathcal{P}}$ due to $G_s$ direction, as \textit{another} upper point. Note that there can be more than one upper point in this version. Also, each time the algorithm picks an upper point, it calculates its $\VP$ and marks the parts of $\mathcal{P}$, as well as $\tau_q$ and $\tau_r$, that fall within the $\VP$. Accordingly, $\RTRV$ can detect the corresponding upper point of each trajectory vertex. It then continues until there are no unmarked trajectory vertices. Once completing this, it starts phase (2) by applying the same procedure used for phase (1). Except that, the algorithm repeats it for every $\VP$ found when running phase (1). Having found the connecting points, $\PNST$ can repeat the idea of $\RTRV$ for every $\VP$ and all distinct pairs of $\VP$s. Yet, the algorithm would need to store every instance of $\RTRV$ separately, which introduces visibility between a pair of $\VP$s. To answer the queries, $\PNST$ can assign a list $l_u$ to each upper point $u$.
    
The algorithm then puts pointers in $l_u$, pointing to instances of $\RTRV$, executed for the corresponding $\VP$ of $u$. Therefore, the algorithm should sort all trajectory vertices and the intersections of the trajectories with $\VP$s based on their distances from the beginning of the trajectory. Finding the predecessor and the successor of a query velocity would then yield the corresponding upper point of the query velocity. Observe that $\PNST$ can determine the intersection points of the trajectories with the boundaries of the $\VP$s, using ray shooting queries.
This can be done when the algorithm sorts the trajectory vertices based on their distances from the beginning of the trajectory. Specifically, picking the last vertex (based on the sorted list) inside a $\VP$ and the first vertex inside the next one. The vertices are adjacent due to the measure chosen for the sorting. The algorithm can then run a ray shooting query using the direction obtained from the vertices it found, and compute the intersection of the trajectory with the $\VP$.

\begin{lemma}
\label{lemma:nonSegPolyTime}
    The $\PNST$ algorithm solves Problem~\ref{TRSP}, subject to the constraints (\ref{constraint2}, \ref{constraint3}) in polynomial time in terms of $n$ and $m$. This algorithm also requires polynomial space in terms of $n$ and $m$. 
\end{lemma}

\begin{proof}
    The correctness and the complexity analysis of the $\PNST$ algorithm (see Sub-subsection~\ref{PNSTAlg}) are as follows:
    
    \textbf{Correctness.} Similar to Theorem~\ref{theorem:RestrictedTRSP}, the heart of $\PNST$ is choosing the connecting vertices. Recall from Remark~\ref{remark:restrictedNonSegRemark} that the aim of finding the connecting vertices is to find \textit{representative} points. Thus, $\PNST$ can determine all visible points on the trajectories. The crucial element, when finding the connecting vertices, is that $\PNST$ computes the $\VP$ of each upper point in phase (1). Also, $\PNST$ marks all the trajectory vertices that fall within the bounds of such a $\VP$. Therefore, there cannot be an area in $\mathcal{P}$ that contains at least one unmarked trajectory vertex. Moreover, since the algorithm uses $G_s$ for picking the vertices, it picks vertices on a \textit{convex chain} of vertices in $V_{q, r}$. The only difference here is that $\PNST$ does not necessarily stop when it picks an upper point. However, by computing each $\VP$, the algorithm does not consider vertices already visible to the upper point it picks. One can see each $\VP$ as a \textit{visibility area} without intersecting the other $\VP$s, having an upper point on its \textit{boundary}. On the other hand, $\PNST$ computes the lower points inside each $\VP$. So, it preserves the locality of the visibility of each $\VP$ and does not pick a lower point s.t. its corresponding upper point might not see it. Again, due to the same argument, the algorithm chooses the lower points in such a way that they become the representative points of the other visibility boundary of their corresponding sub-trajectories.
    
    \textbf{Time and Space Complexity.} A set of reflex vertices $V_R \subset \mathcal{P}$ can exist s.t. each $r_{\mathcal{P}} \in V_R$ blocks the visibility of $\tau_q$ and/or $\tau_r$ every time the trajectory(s) goes \textit{behind} $r_{\mathcal{P}}$. Recall from the algorithm presented in Subsection~\ref{nonSegmentRestricted} that it uses $G_S$. Thus, $\PNST$ keeps tracing the vertices on a \textit{convex chain} (as defined in Subsection~\ref{subsec::Defs}) of vertices in $V_{q, r}$. So, depending on the size of $V_R$, the algorithm would then pick more upper points. Note that if there are a constant number of such reflex vertices, $\PNST$ picks one or a constant number of upper points, similar to \ref{nonSegmentRestricted}. However, in the worst case, $|V_R| \in \mathcal{O}(n)$ s.t. $n$ is the number of vertices in $\mathcal{P}$. Denote the overall number of trajectory vertices as $m$. Accordingly, there can be $\mathcal{O}(\frac{m}{n})$ trajectory vertices inside each $\VP$ s.t. the number of vertices in each $\VP$ gets maximized. So, the cost to repeat the idea of using \textit{half-plane queries} becomes proportional to $\mathcal{O}(mn)$.
\end{proof}

\subsubsection{The $\INST$ Algorithm}
\label{INSTAlg}

The time complexity of $\PNST$ can be improved in $\INST$, by adding an extra step, as follows: For every $r \in V_R$, compute a $\VP$, bounded by the extension of the line segment between $r^{'} \in V_R$ and $r$ s.t. $r^{'}$ is the \textit{next} reflex vertex from $r$, in terms of the $G_s$ direction. Note that a ray-shooting query is required in both directions of $\overline{r r^{'}}$ to find the intersection points with $\mathcal{P}$. While computing every $\VP$, mark all trajectory vertices met, to avoid considering them in $\VP$s that $\INST$ might find in the future. Moreover, mark all reflex vertices like $r^{'}$ in $V_R$ that fall within the $\VP$ of $r$. Once computed the $\VP$ from $r$, run $\PNST$ and then repeat the whole procedure for the rest of the reflex vertices that are still unmarked. Note that $\INST$ bounds the execution of $\PNST$ to every $\VP$ it finds. Thus, the algorithm would not reach more reflex vertices than the ones in $V_R$. Note that such reflex vertices force finding several upper points on the vertices of only one of the trajectories. Observe that $\PNST$ had to check whether there are trajectory vertices inside the corresponding $\VP$ of each upper point like $u$ (say ${\VP}_u$) s.t. their visibility is not blocked by reflex vertices like $r$. However, $\INST$ computes a $\VP$ from $r$ and marks all reflex vertices inside the $\VP$ of $r$. The algorithm then repeats the procedure for the unmarked reflex vertices until marking all of them.

\begin{theorem}
\label{theorem:INST}
    The $\INST$ algorithm offers $\mathcal{O}(n \log n + m(\log n + \log m))$ running time to solve Problem~\ref{TRSP}, requiring $\mathcal{O}(n + m)$ space and $\mathcal{O}(n)$ time for pre-processing. This algorithm is subject to constraints (\ref{constraint2}, \ref{constraint3}). It also provides $\mathcal{O}(\log m + k)$ query time.
\end{theorem}
\begin{proof}

        The correctness and the complexity analysis of the $\INST$ (see Sub-subsection~\ref{INSTAlg}) are as follows:

        \textbf{Time and Space Complexity.} There will be no need to check the visibility between every pair of $\VP$s and pay the $\mathcal{O}(mn)$ cost. So, $\INST$ would naturally find out if the visibility is not blocked for some trajectory vertices in ${\VP}_u$. Therefore, there is no need to perform the $\mathcal{O}(mn)$ operation. Observe that $\INST$ no longer requires running $\mathcal{O}(mn)$ instances of $\PNST$. Accordingly, the memory consumption remains $\mathcal{O}(n + m)$. Note that all reflex vertices inside this area will be marked when combining the $\VP$ constructed for $r$ and its corresponding connecting vertex. So, $\VG$s will not overlap in terms of polygon and trajectory vertices.

        \textbf{Correctness.} W.lo.g consider $\nu \in \tau_q$ a point (a trajectory vertex or any other point) visible for at least one point like $x$ on $\tau_r$. Denote the data structure $\INST$ finally generates as $I$. Let us assume that $\nu$ will not be reported if we query $I$ with velocity $v_r$, which brings the moving entity $r$ to $x$ according to Definition~\ref{timestampVelocity}. Recall from Lemma~\ref{lemma:nonSegPolyTime} that $\PNST$ preserves the locality of the visibility of each $\VP$ and does not pick a lower point s.t. its corresponding upper point might not see it. Also, notice that $\INST$ picks $r \in V_R$ and computes a $\VP$, bounded by the extension of the line segment between $r^{'} \in V_R$ and $r$ s.t. $r^{'}$ is the \textit{next} reflex vertex from $r$, in terms of the $G_s$ direction. Accordingly, $\INST$ considers all trajectory vertices that \textit{could be visible} to the upper point it picked if $r$ did not exist. Therefore, w.l.o.g if there are trajectory vertices not visible for the upper point $u \in \tau_q$, yet visible for at least one trajectory vertex $\nu^{'} \in \tau_r$ s.t. $\nu^{'} \in \VP$, then $\INST$ detects it. The rest of the procedure that focuses on each $\VP$ will allow reporting $\nu$ due to Lemma~\ref{lemma:nonSegPolyTime}. Accordingly, a point like $\nu$ does not exist. 
        
        Also, see Figure~\ref{exampleGS} for further illustrations.
\end{proof}

\begin{observation}
\label{optObs}
    The $\INST$ algorithm offers the optimal query time under the settings of $\TRVP$ appeared in Theorem~\ref{theorem:INST}.
\end{observation}
\begin{proof}
    The data structure offered by Chazelle et al. has an optimal query time~\cite{chazelle1985power}, answering the half-plane query in $\mathcal{O}(\log p + k)$ s.t. $p$ is the total number of points and $k$ is the size of the result reported. Denote this data structure as $H$. Consider the settings of $\TRVP$ in Theorem~\ref{theorem:INST}. Accordingly, $\tau_q$ and $\tau_r$ are the corresponding trajectories of the moving entities $q$ and $r$, respectively. Also, let $P$ be the set of trajectory vertices and intersections of bi-tangents with the trajectories, determined based on an arbitrary $\VG$ (as discussed in Theorem~\ref{theorem:INST}). Moreover,  let $A_P$ be the area in which all points in $P$ exist. Therefore, $H$ can process $P$ and consider a half-plane as the boundary of the visibility for an arbitrary point like $x$ on $\tau_q$ (or $\tau_r$). Note that $x$ is within $A_P$. Observe that $H$ can find the points falling into the visible area of $x$ by performing the half-plan query. In addition, there are always some points in $P$ visible to $x$, because otherwise whether $\VG$ becomes empty or $x$ will not fall within $A_P$. On the other hand, recall that $H$ is the optimal data structure for the half-plan query. Also, finding the visible points in $P$ for $x$ is equivalent to performing the half-plane query. So, the data structure suggested in Theorem~\ref{theorem:INST} is optimal for the $\TRVP$.
\end{proof}


\section{Discussion}
We extended the previous works by solving the general version of the Trajectory Visibility problem. In particular, for a constant query velocity of a moving entity, we specify all visible parts of the other entity's trajectory and every possible velocity of the other entity to become visible.

There are possible directions for future improvements to our current results. One can find an efficient way of supporting curve trajectories. It is also interesting to focus on the cases of polygons with holes.

\bibliography{lipics-v2021-sample-article}

\appendix

\section{Visibility Glass}
\label{VGAppendix}

Referring to~\cite{guibas1989optimal}, define for all two line-segments $s_1$ and $s_2$ in a simple polygon $\mathcal{P}$, the hourglass $H(s_1, s_2)$ to be the union of all shortest paths between points on $s_1$ and $s_2$. The hourglass $H(s_1, s_2)$ is a subset of $\mathcal{P}$, bounded by  $s_1$ and $s_2$, as well as the two shortest paths between their endpoints. More specifically, denote the first endpoint of $s_1$ as $P^{+}_{s_1}$ and the second one as $P^{-}_{s_2}$. Similarly, define $P^{+}_{s_2}$ and $P^{-}_{s_2}$ for $s_2$. The upper chain, as shown in Figure~\ref{Fig2}, is thus the shortest path between $P^{+}_{s_1}$ and $P^{+}_{s_2}$, and the lower chain is the shortest path between $P^{-}_{s_1}$ and $P^{-}_{s_2}$. In case $s_1$ and $s_2$ are not vertical yet their corresponding endpoints lie in a convex position, one can rotate the plane until one of the line segments becomes vertical. On the other hand, if the endpoints do not lie in a convex position, then the upper and the lower chain would share an endpoint, which is a simpler case. Accordingly, denote the visibility glass~\cite{eades2020trajectory} as $L(s_1, s_2)$ the (possibly empty) union of line segments between $s_1$ and $s_2$ that are contained in $\mathcal{P}$. Observe that either $L(s_1, s_2)$ is empty or there exist sub-line-segments $s^{'}_{1} \subset s_1$ and $s^{'}_{2} \subset s_2$ s.t. $L(s_1, s_2) = H(s^{'}_{1}, s^{'}_{2})$. Moreover, $s^{'}_{1}$ and $s^{'}_{2}$ are bounded by two bi-tangents on the shortest paths between the endpoints of $s_1$ and $s_2$ (E.g. see Figure~\ref{Fig1}). One last crucial element proved in~\cite{guibas1987linear}, is that the shortest path between two points in $\mathcal{P}$ can be computed in linear time. So, the overall running time of constructing $L(s_1, s_2)$ is linear in terms of the number of vertices of $\mathcal{P}$.

\section{Proof of Lemma~\ref{lemma:VRT}}\label{proof:lemmaVRT}

\textit{Lemma~\ref{lemma:VRT}}
    \textit{The Visibility Range Tree Algrithm($\VRT$) solves Problem~\ref{TRGV} with the error rate of $\frac{1}{L} \leq \varepsilon$, offering $\mathcal{O}(t \log t \log n)$ running time, where $t = \max \{ {len}(\tau^{*}_{q}) \cdot L, \ {len}(\tau^{*}_{r}) \cdot L \}$. Given a query small segment with length $\frac{1}{L}$ and/or a \textit{range} (i.e. a set of contiguous small segments), the algorithm offers $\mathcal{O}(\log t  + k)$ time to report all timestamps that the query segment or range becomes visible s.t. $k$ is the number of such timestamps. The space consumption of the algorithm is $\mathcal{O}(t \log t)$ and it requires $\mathcal{O}(n)$ pre-processing time (see Lemma~\ref{lemma:completeVis}).}

    \begin{proof}
       The complexity analysis of $\VRT$ (see Sub-subsection~\ref{VTRAlg}) is as follows:
    
        \textbf{Space Complexity}. Since $\VRT$ assumes $\mathcal{P}$ as a simple polygon without holes, the space required for the ray shooting algorithm remains $\mathcal{O}(n)$. Also, since $\VRT$ builds a range tree, the space consumption will not exceed the bound mentioned in Lemma~\ref{lemma:VRT}.
    
        \textbf{Time Complexity}. The ray shooting algorithm requires a pre-processing time of $\mathcal{O}(n)$. Also, each ray shooting operation takes $\mathcal{O}(\log n)$ time, in which $n$ is the number of vertices in $\mathcal{P}$ \cite{chen2015visibility}. It is clear that the running time of finding $e_m$ is bounded by $\mathcal{O}(\log n \cdot \log (\frac{d}{\varepsilon}))$ i.e. $\mathcal{O}(\log n \cdot \log \frac{1}{\varepsilon L})$. The $\mathcal{O}(\log n)$ term is because $\VRT$ requires a ray-shooting query each time the visibility of $e_m$ gets checked. However, the length of a \textit{small} segment is $d = \frac{1}{L}$. So, choosing $\varepsilon$ carefully to be $\frac{1}{L} \leq \varepsilon$, nullifies the need for performing a binary search operation.
    
        Note that due to the definitions and for a segment $s$, there are $\mathcal{O}(t)$ time intervals overall s.t. $t = \max \{ {len}(\tau^{*}_{q}) \cdot L, \ {len}(\tau^{*}_{r}) \cdot L \}$. On the other hand, since each timestamp might appear in at most $\mathcal{O}(\log t)$ range tree nodes, the overall cost to search all timestamps can be bounded by $\mathcal{O}(t \log t)$. Also, note that since timestamps are naturally sorted, there is no need to pay an extra cost in $\nu$ to sort them. The algorithm only has to search on the list in $\nu$ and discard the timestamps older than the query timestamp. Accordingly, it becomes now rudimentary to specify the range(s) visible to a viewer at each timestamp. Also, finding the \textit{time interval}(s) during which a \textit{small segment} and/or a \textit{range}\footnote{A set of contiguous small segments} is visible to the viewer, gets trivial as well.
    \end{proof}
    
\section{Proof of Theorem~\ref{theorem:RestrictedTRSP}}\label{proof:theoremRestrictedTRSP}

\textit{Theorem~\ref{theorem:RestrictedTRSP}}\textbf{.}
    \textit{ The $\RTRV$ algorithm provides $\mathcal{O}(n + m (\log m + \log n))$ running time to solve Problem~\ref{TRSP}, subject to the constraints (\ref{constraint2},  \ref{constraint3}), and Remark~\ref{remark:restrictedNonSegRemark}. The algorithm requires $\mathcal{O}(n)$ pre-processing time and only $\mathcal{O}(n + m)$ space. The query time is bounded by $\mathcal{O}(\log m + k)$.}

    \begin{proof}

        The correctness and the complexity analysis of $\RTRV$ algorithm (see Sub-subsection~\ref{RTVAlg}) are as follows:

        \setlength{\textfloatsep}{10pt}
        \setlength{\intextsep}{10pt}
        \begin{figure*}[htp]
            \centering
            \includegraphics[scale=.7]{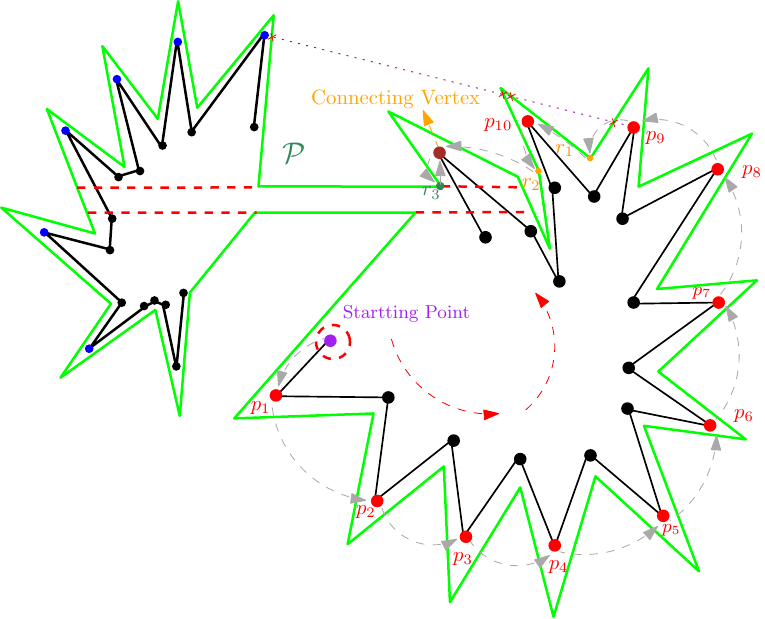}
             \caption{In this example, $G_s$ starts from \textit{starting point}. Then it continues the scan until reaching $p_9$ whose visibility on the next vertex (due to $G_s$'s order) is blocked. So it jumps on the reflex vertex $r_1$ from which $G_s$'s order shows $p_{10}$. The $G_s$ then similarly continues until reaching $r_3$. Yet, as $r_3$ has visibility on the other trajectory, the connecting vertex would be specified.}
            \label{exampleNonSegTraj}
        \end{figure*}

        \textbf{Correctness.} The key part in $\RTRV$ is finding the connecting vertices. Accordingly, it seems necessary to provide a separate proof for its correctness: Observe that only \underline{reflex vertices} cause a block of visibility when searching for connecting vertices. Now suppose that there is a sub-trajectory in $\tau_q$, say $\tau^{*}_{q}$, that sees another sub-trajectory in $\tau_r$, say $\tau^{*}_{r}$. Assume that $\RTRV$ \textit{does not} specify connecting points in such a way that $L(\overline{v^{1}_q v^{2}_q}, \overline{v^{1}_r, v^{2}_r})$ will cover $L(\tau^{*}_{q}, \tau^{*}_{r})$. Referring to \ref{nonSegmentRestricted}, recall that $\RTRV$ starts $G_s$ on the union of all trajectory vertices in $\tau_q$ and $\tau_r$, named $V_{q, r}$. W.l.o.g, suppose the algorithm starts picking vertices from $\tau_q$. While $\RTRV$ keeps picking, it can rely on $G_s$ for picking the vertices as a \textit{convex chain} (as defined in Subsection~\ref{subsec::Defs}) of the vertices in $V_{q, r}$ (for instance, see Figure~\ref{exampleNonSegTraj}). When $\RTRV$ finally reaches a vertex $\nu_q \in \tau_q$ from which $G_s$ picks a vertex $\nu_r \in \tau_r$, there can be two cases: First, if $\nu_r$ is visible to $\nu_q$, $\RTRV$ still preserves the same property it relied on while picking vertices from $\tau_q$ (see Figure~\ref{exampleGS} - a). So, there cannot be vertices on the trajectories that can form $\tau^{*}_{q}$ and $\tau^{*}_{r}$. Second, if $\RTRV$ continues $G_s$ from a reflex vertex on $\mathcal{P}$ (see Figure~\ref{exampleGS} - b) and picks a vertex on $\tau_r$, it can be observed that there still cannot be vertices so that $\tau^{*}_{q}$ and $\tau^{*}_{r}$ exist. Because, while there is a reflex vertex that blocks the visibility of some vertices in $\tau_q$, there can only be $\tau^{*}_{q}$ and $\tau^{*}_{r}$ in the areas that are still visible. Also, when $\RTRV$ picks a vertex from $\tau_r$, it chooses the vertex on $\tau_q$ met before the reflex vertex as a connecting vertex. Therefore, there still cannot be trajectory vertices that might create $\tau^{*}_{q}$ and $\tau^{*}_{r}$. 

        \setlength{\textfloatsep}{10pt}
        \setlength{\intextsep}{10pt}
        \begin{figure*}[htp]
            \centering
            \includegraphics[scale=.65]{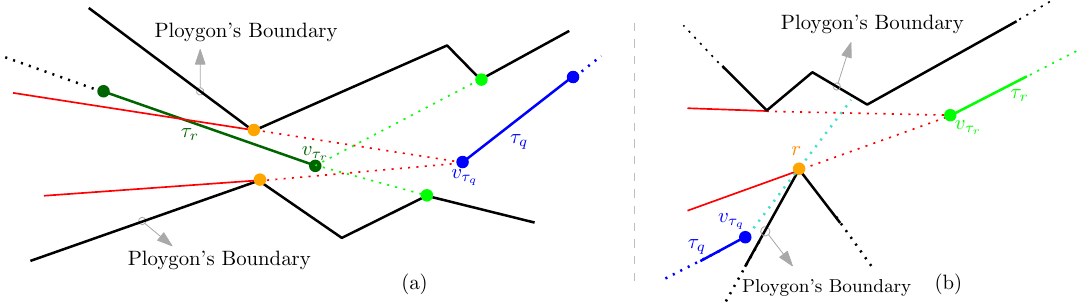}
             \caption{$(a)$ As this figure illustrates, $G_s$ finds vertices on $\tau_q$ and $\tau_r$ that have sight on one another. So, it marks them as $v_{\tau_r}$ and $v_{\tau_q}$. $(b)$ In this case, $G_s$ hits the reflex vertex $r$, while $r$ has visibility on the other trajectory. Thus, it picks the connecting vertices $v_{\tau_r}$ and $v_{\tau_q}$.}
            \label{exampleGS}
        \end{figure*}

        \textbf{Time and Space Complexity.} $\RTRV$ requires storing the vertices of trajectories $\tau_q$ and $\tau_r$, and the vertices of the polygon $\mathcal{P}$. Also, the algorithm reuses the method discussed for $\LRTV$ by constructing $\VG$. This implies the algorithm consumes $\mathcal{O}(n + m)$ in terms of memory. On the other hand, the algorithm hires half-plane queries to find the velocity ranges.  Such queries take $\mathcal{O}(\log m  + k)$ time~\cite{chazelle1985power} to detect such points, in which $m$ is the overall number of trajectory vertices in $\tau_q$ and $\tau_r$. Also, $k$ is the number of vertices and intersections that fall within the visible area. According to Lemma~\ref{lemma:completeVis}, the algorithm needs $\mathcal{O}(m \log n)$ execution time to check the complete visibility of the trajectories. Also, performing half-plane queries requires processing the trajectory vertices in $\mathcal{O}(m \log m)$ time. Note that we consider this as a step of the algorithm and not the pre-processing phase. However, observe that pre-processing is needed only for the ray-shooting queries, taking $\mathcal{O}(n)$ of running time.
    \end{proof}
\section{Notations}
\label{sec:notations}
\begin{itemize}
    \item $n$ = the number of the vertices of the given simple polygon $\P$.
    \item $\tau$ = a trajectory.
    \item $m$ = the summation of the vertices of both trajectories.
    \item $k$ = the number of velocity ranges reported in the output for the case of having non-constant complexity trajectories.
    \item $T_{q, r} = $ the set of time intervals at which $q$ and $r$ can see each other.
    \item $[t_i, t_j] = $ a member of $T_{q,r} \ {s.t.} \ j > i \ \& \ i, j \in \mathbb{N}$.
    \item $L(\tau_q, \tau_r) = $ the visibility glass of $\tau_q$ and $\tau_r$.
    \item $P^{+}_{s} = $ the \textit{upper} endpoint of an arbitrary line segment $s$. The word \textit{upper} is merely a naming convention. 
    \item $P^{-}_{s} = $ the \textit{lower} endpoint of an arbitrary line segment $s$. The word \textit{lower} is merely a naming convention.
    \item $G_s = $ the Graham Scan procedure. 
\end{itemize}
\end{document}